\documentclass[singlecolumn]{revtex4-2}
\usepackage{amssymb}
\usepackage{xcolor}
\usepackage{hyperref}
\usepackage{amsmath}
\usepackage{mathtools}
\usepackage[title,toc]{appendix}
\usepackage[a4paper, total={6in, 9in}]{geometry}
\usepackage[normalem]{ulem}
\usepackage{caption}
\usepackage{subcaption}
\newcommand{\floor}[1]{\left\lfloor #1 \right\rfloor}
\usepackage{tikz}
\usetikzlibrary{shapes.geometric, arrows,positioning}
\tikzstyle{startstop} = [rectangle, rounded corners, minimum width=3cm, minimum height=1cm,text centered, draw=black, fill=red!30]
\tikzstyle{io} = [trapezium, trapezium left angle=70, trapezium right angle=110, minimum width=3cm, minimum height=1cm, text centered, draw=black, fill=blue!30]
\tikzstyle{process} = [rectangle, minimum width=3cm, minimum height=1cm, text centered, draw=black, fill=orange!30]
\tikzstyle{decision} = [diamond, minimum width=3cm, minimum height=0.5cm, aspect=2.5,text centered, draw=black, fill=green!30]
\tikzstyle{arrow} = [thick,->,>=stealth]

\usepackage{rotating}

\begin{document}

\title{Order Parameter Engineering for Random Systems}
\author{G. Anand}
\email[]{ganand@metal.iiests.ac.in}
\affiliation{
	Department of Metallurgy and Materials Engineering, Indian Institute of Engineering Science and Technology, Shibpur, Howrah, WB, India 711103 }
\author{Swarnava Ghosh}
\affiliation{
Centre for Computational Sciences, Oak Ridge National Laboratory, 5200, 1 Bethel Valley
Rd, Oak Ridge, 37830, TN, USA}
\author{Markus Eisenbach}
\affiliation{
Centre for Computational Sciences, Oak Ridge National Laboratory, 5200, 1 Bethel Valley
Rd, Oak Ridge, 37830, TN, USA}
\begin{abstract}
	The chemical short-range order (CSRO) in the crystalline materials influences the properties and its effect is particularly important in the context of the multicomponent materials. We propose a scheme for CSRO parameter or $\mathrm{\Delta - parameter}$ in terms of number of like and unlike bonds in the multicomponent systems. The \textbf{OPERA} or \textbf{\underline{O}}rder \textbf{\underline{P}}arameter \textbf{\underline{E}}ngineering for \textbf{\underline{RA}}ndom Systems scheme for both semi-canonical as well as canonical ensemble is proposed. The proposed framework provides a high-throughput scheme for exploration of the CSRO in terms of the $\mathrm{\Delta}$-parameter. We demonstrate the applicability of the $\mathrm{\Delta}$-parameter for describing the CSRO in multicomponent alloys and oxides (FCC-CoCrNi, BCC-MoNbTaW and (CoCuMgNiZn)O). We show that the $\mathrm{\Delta}$-parameter not only inherits the merit of Warren-Cowley order parameter, but it also addresses its limitations. The $\mathrm{\Delta}$-parameter being a scalar quantity can represent the chemical order and OPERA framework can deal with both semi-canonical and canonical ensemble.
\footnote{\tiny{This manuscript has been authored in part by UT-Battelle, LLC, under contract DE-AC05-
00OR22725 with the US Department of Energy (DOE). The US government retains and the publisher, by accepting the article for publication, acknowledges that the US government retains a nonexclusive, paid-up, irrevocable, worldwide license to publish or reproduce the published form of this manuscript, or allow others to do so, for US government purposes. DOE will provide public access to these results of federally sponsored research in accordance with the DOE Public Access Plan (http://energy.gov/downloads/doe-public-access-plan).}}		
\end{abstract}

\maketitle
\section{Introduction}
The chemical short range order (CSRO) has garnered significant attention in recent times due to their direct influence on the properties. Such interest has been particularly prominent in high-entropy materials community. Initially, high-entropy alloys were assumed to be perfectly chemically-disordered with elements residing randomly on the lattice sites. However, such a view was changed with both the experimental \cite{santodonato2015deviation,zhang2020short} as well as theoretical investigations \cite{Widom2014,Fernandez2017,Singh2015}, which pointed that CSRO is a critical structural information, which needs to be considered to develop the understanding about structure-property correlation in high-entropy materials. The CSRO can influence the mechanical properties of high-entropy alloys \cite{Wang2021,Antillon2020,zhang2019direct,zhang2017local}. It has been demonstrated that the local-chemical ordering can influence the dislocation activity in high-entropy alloys \cite{Sudmanns2021}. The CSRO can not only influence the Frank-Read source activation \cite{Smith2020}, but it also influences the cross-slip of the screw dislocations by increasing the activation barrier for dislocation movement leading to the reduction in the dynamic recovery \cite{Abu2022}. It has been documented that CSRO is influenced by lattice distortion \cite{He2021}. The CSRO also tunes the melting behaviour of HEA \cite{Jian2021}, localizes diffusion \cite{Xing2022} and influences defect evolution in HEA\cite{Zhao2021}. In CoCrNi, the CSRO has been reported to be magnetically driven with repulsion between Co-Cr and Cr-Cr bond pairs \cite{Walsh2021}. CSRO not only influences properties, but it can also act as nucleation sites for BCC to HCP phase transformation in refractory high-entropy alloys. Presence of the multi-hyperuniform long-range order in high and medium entropy semiconductor alloys lead to the emergence of CSRO, which further decreases the configurational energy of the configurations, in comparison to the structures without CSRO. It has been shown that such CSRO may be the reason behind the successful application of rule-of-mixture approach for the prediction of properties in multi-component semiconductor alloys such as \cite{Chen2021}. The CSRO has been shown to not only influence the structural properties in the context of alloys, but it plays crucial role in tuning the functional properties of other kind of high-entropy materials, such as disordered rocksalt-structured materials \cite{cai2022thermodynamically} and high-entropy rare-earth niobates and tantalates \cite{wright2022short}.\\       
The CSRO is a local atomic property and it is not straightforward to determine such property through experimental approaches, such as high-resolution transmission electron microscopy studies \cite{zhang2022characterization,zhou2022atomic,chen2021direct,zhang2020short}, neutron total scattering with extended x-ray absorption fine structure\cite{zhang2017local,owen2016new} technique, etc. In order to deal with such a challenge, computational approaches involving Monte Carlo simulation with either \emph{ab-initio} calculations \cite{xian2020prediction,tamm2015atomic}, cluster-expansion formalism \cite{sobieraj2020chemical,Eisenbach2019,Fernandez2017} or with interatomic potential (classical \cite{shen2021kinetic,maiti2016structural,huang2021atomistic} or machine-learned \cite{byggmastar2021modeling} interatomic potentials) has been carried out. Hybrid Monte Carlo-Molecular Dynamics approach, in which the molecular dynamics can mimic the finite temperature oscillations of  atoms at their lattice sites, while Monte Carlo swaps of atomic species between lattice sites are carried out. Such approach could predict the CSRO in MoNbTaW  high-entropy alloy.\cite{Widom2014,Antillon2020}. Another approach involves the application of the linear-response theory for the prediction of CSRO in HEA \cite{Singh2015}. In such an approach, electronic structure calculations are employed to explain CSRO at high temperatures and incipient long-range order in high entropy alloys. In above-stated approaches, the emphasis is on the identification of CSRO through tedious experimental approaches or through computationally expensive methods. However, we aim to address the issue of CSRO through inverse methodology. In the present approach, we aim to develop a method for generating CSRO in the atomic configuration and studying the effect of CSRO on the properties in contrast to the traditional approaches, where the CSRO needs to be determined first and correlation between CSRO and properties are ultimately developed. In the present approach, CSRO for any crystalline material can be be tuned in the supercell approach, which may be later used for simulating the variation in the properties. It should be noted that, Yu \emph{et. al.} have also developed random structure generator with short-range order which employs Markov-Chain Monte Carlo method \cite{Yu2019}. In addition to the above, we aim to develop the CSRO generation scheme, which does not involve the energy calculation step and hence numerous supercells with desired CSRO can be generated aiding the high-throughput computational exploration of CSRO in chemically-disordered crystalline lattice. In addition to the above, it should be noted that the proposed approach can also be employed to generate the structures, which are far-from-equilibrium. Such configurations can be particularly helpful for studying CSRO phenomena through irradiation \cite{su2022radiation,zhang2017local} or deformation induced CSRO \cite{song2022room}, which might not be accessible through the traditional approaches, which can only generate the equilibrium ordered structure. The present approach can also be employed for the development of multipoint ordering in complex multicomponent materials \cite{goff2021quantifying}. This approach can also be used for generation of supercell with desired CSRO, which at this stage need to be either determined through computationally expensive approaches, as discussed earlier.     

\section{Defining the $\mathrm{\Delta-parameter}$}
The concept of order parameter stems out of Landau's critical transition theory \cite{callen1998thermodynamics}. The motivation to define order parameter was to quantify any kind of phase transition. Landau predicted that order parameter should attain zero value in the disordered phase, while it should have a non-zero value in the ordered low temperature phase. In addition to the above, such order parameter should have different values in different phases. The theory of order parameters was also generalized to determine the critical point in phase transition, which is quantified in terms of divergence of susceptibility of the order parameter. However, for the configurational order parameter, such susceptibility is undefined. Hence, configurational order parameter should have zero value at complete disorder and non-zero value due to ordering and it should have different values for different crystal structures. Initial attempt to describe the ordering and order-disorder transition conceptually addresses the problem as disorder arising due to thermal agitation. In such cases, the long-range order parameter ($\mathrm{S}$) is defined as fraction, which is $\mathrm{0\; \leq \;S\; \leq\; 1}$, where the value of $\mathrm{0}$ quantifies complete disorder, while $\mathrm{1}$ implies the complete order. Note that in such a framework, S is defined to be $\mathrm{S = \frac{p-r}{1-r}}$, where, p is the probability of finding atom in its `right' position. Note that by atom in its `right' position, it's implied that in the ordered state a particular atom is residing in its sub-lattice. While, `r' denotes the the fraction of sites, which is occupied by this particular atom. \\
In Bragg-Williams (BW) approximation \cite{bragg1935effect}, the energy change due to the movement of the system from the perfect order (which is assumed to be of minimum) to the perfect disorder involves the change in the energy. Such change in energy arises due to interchange of atoms from their `right' to `wrong' positions. In this approach the drop of order-parameter to zero value has been rationalised in terms of energy required for interchanging the atomic position in the course of disordering. It has been considered that the major interaction between atoms arise due to the nearest-neighbour interaction and hence, the concept of short-range order was introduced, which in this approach could be calculated using nearest-neighbour pair interaction energy. Formally, short-range order in BW approach is defined in the term of difference in the probability of finding a unlike and like atoms. As the disorder increases, the number of `wrong' pairs (like atoms) increases and at the complete disorder, the number of `right' and 'wrong' pairs are equal. It should be noted that in the BW approach, two opposing tendency of ordering, \emph{i.e.,} superlattice formation and clustering are being considered to yield a scalar value to quantify the short-range order. In such a scenario, when the temperature is lower than the order-disorder transformation temperature, then the short-range order is an intermediate stage of the long-range order. \\
The Warren-Cowley (WC) order parameter \cite{cowley1965short} has been extensively applied for describing the ordering behaviour. Here, the order parameter for the species $\mathrm{i}$ and $\mathrm{j}$ may be expressed in term of ($\mathrm{\alpha_{ij}}$), which is 
\begin{equation}\label{eq:wc}
    \centering
    \mathrm{
    \alpha_{ij} = 1 - \frac{P(j|i)}{c_j}
    }
\end{equation}
where $\mathrm{P(j|i)}$ is the probability of finding $\mathrm{j}$ atom near $\mathrm{i}$ atom and $\mathrm{c_j}$ is the concentration of $\mathrm{j}$ atom. The Warren-Cowley order parameter can have value between $\mathrm{-1}$ and $\mathrm{1}$, with value of $\mathrm{-1}$ implying complete superlattice formation, while at $\mathrm{1}$ complete clustering exists, and zero value describing complete disorder. The WC order parameter needs to be defined for each pair. This order parameter provides a way to correlate the chemical ordering with the peak of autocorrelation or Patterson function. Note that the peak positions of the autocorrelation function are dependent on the interatomic distances. Hence, WC order parameter provides the information about the average environment of the atom. It should be noted that $\mathrm{\alpha_{ij}}$ can describe the intensity of diffused scattering in the reciprocal lattice, and therefore these order-parameters have direct experimental validity \cite{cowley1965short}. \\
Warren-Cowley order parameter were originally developed for binary alloys with Bethe's approximation or pair approximation to deal with ordering. The Warren-Cowley approach has been extended to multiple-component high entropy alloys. However, High-entropy materials or multiple-component materials pose a significant challenge in accurately describing ordering. For example, the Warren-Cowley approach cannot describe the ordering motif, where higher-point ordering than two-point ordering are possible (for \emph{e.g.,} oxides with multiple metal atoms at cation sites) \cite{goff2021quantifying}. Also, instead of one scalar value representing the degree of randomness, it is represented as matrix having order parameter for each of the atomic pairs. To overcome these issues of the Warren-Cowley approach in describing ordering in High-Entropy materials, there has been recent efforts to quantify the order parameter as a single scalar value, capable of describe the ordering in chemically complex structure \cite{He2021,yin2021neural}. \\
In view of above, we describe an order parameter for chemical ordering in High-Entropy materials. In our approach, the degree of disorder is expressed in terms of a scalar value, capable of dealing with chemically complex materials, while keeping the merits of the Warren-Cowley order parameter. Furthermore, such a scalar order parameter, which can distinguish between like and unlike atom ordering does not need the explicit energy calculations.  
The short-range order parameter (SRO) is defined in terms of the $\mathrm{\Delta}$ parameter, which may be expressed as:
\begin{equation}\label{one_eq}
\mathrm{
    \Delta = \beta_{i,j}+\beta_{i,i}
    = \frac{\sum\limits_{i \neq j}^{K_1}{\left(1-\frac{m_{ij}}{2n}\right)}}{K_{1}} + \frac{\sum\limits_{i = j}^{K_2}{\left(\frac{m_{ij}}{n} - 1\right)}}{K_2}
    }
\end{equation}
In equation \ref{one_eq}, $\mathrm{\beta_{ij}}$ and $\mathrm{\beta_{ii}}$ are the components of $\mathrm{\Delta-parameter}$ for unlike and like bonds, respectively, $\mathrm{n}$ is the number of like-bonds in the perfectly disordered solid-solution, $\mathrm{K_1}$ and $\mathrm{K_2}$ represent the number of type of unlike and like bonds, respectively and $\mathrm{m_{ij}}$ and $\mathrm{m_{ii}}$ represent the number of unlike and like bonds in the supercell, respectively. We note that equation \ref{one_eq} has been derived with an understanding that $\Delta$ parameter may vary between $\mathrm[-1,1]$ with attaining the value of $\mathrm{0}$ at the complete chemical disorder. The number of like-bonds $\mathrm{n}$ in the perfectly disordered solid-solution can be expressed as  
\begin{equation}\label{two_eq}
\mathrm{
    n=\frac{N}{(2K_{1}+K_{2})}
    }
\end{equation}
Note that the probability of finding the unlike bonds is twice the like-bonds in the solid-solution (see section-1 in the supplementary information (SI)) and this is the reason for the $\mathrm{2n}$ as normalising factor in the $\mathrm{\alpha_{ij}}$ and $\mathrm{2K_1}$ in the denominator of equation \ref{two_eq}. The $\mathrm{N}$ is the above equation is the total number of bonds in the supercell. Note that there is no cutoff in the $\mathrm{\Delta-parameter}$, so the determination of $\mathrm{m_{ij}}$  is not limited to the identity of the coordination shell, rather it is being calculated for the full supercell, making this parameter capable of dealing with multi-point ordering.\\
The introduction of the CSRO in the supercell can be thought of as deviation from the disorder in the supercell \cite{rao2022analytical}, where the positive value of $\mathrm{\Delta-parameter}$ can be attained by increasing instances of like-bond(s) and negative value of $\mathrm{\Delta}$-parameter can be attained by increasing the instances of unlike bond(s). Since, there are $\mathrm{K_1}$ types of unlike bonds and $\mathrm{K_2}$ types of like bonds, the particular value of $\mathrm{\Delta}$ can be imposed by increasing the propensity of a particular bond-type. So, the first step of the generation of supercell with desired CSRO involves the generation of chemically disordered structure.

\section*{Random structure generation}
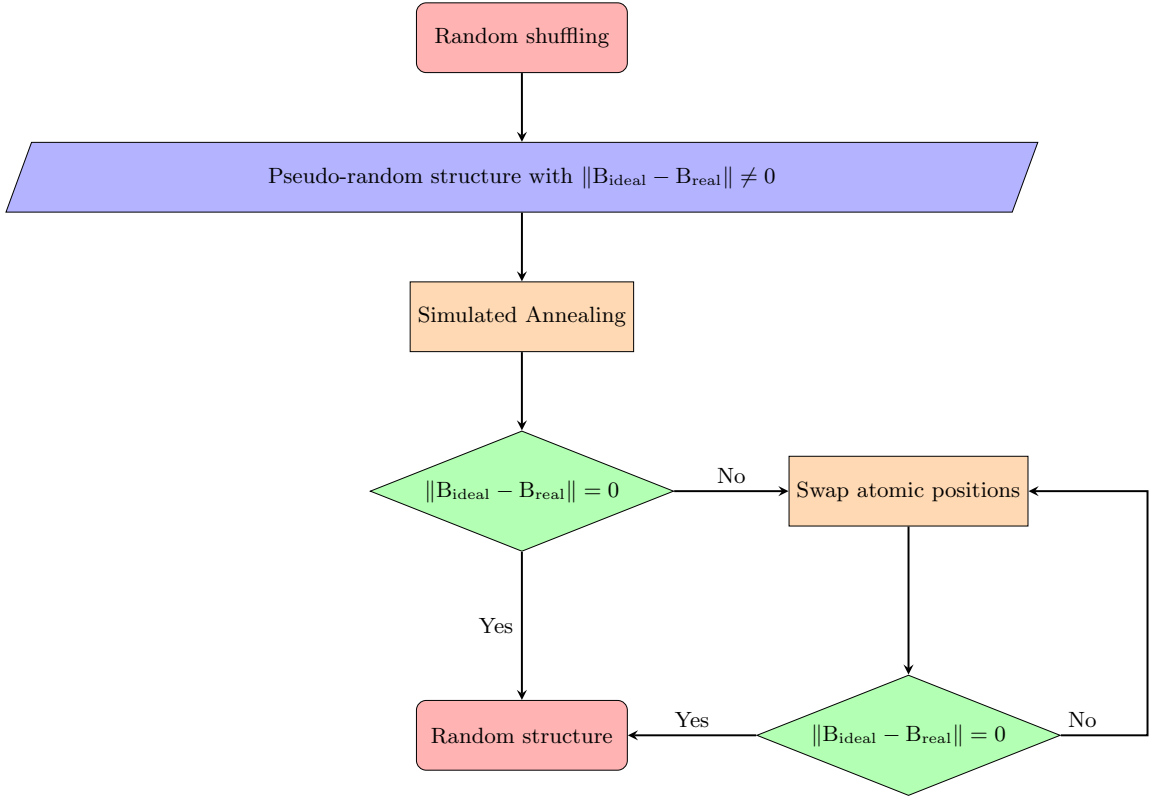
\begin{figure}
\centering
\resizebox{\textwidth}{!}{
\begin{tikzpicture}[node distance=2cm]
\centering
\node (start)[startstop]{Random shuffling};
\node (in1) [io, below of=start] {Pseudo-random structure with $\mathrm{\Vert B_{ideal} - B_{real} \Vert} \neq 0$};
\node (pro1) [process, below of=in1] {Simulated Annealing};
\node (dec1) [decision, below of=pro1,yshift=-0.5cm] {$\mathrm{\Vert B_{ideal} - B_{real} \Vert} = 0$};
\node (pro2a) [startstop, below of=dec1, yshift=-1.5cm] {Random structure};
\node (pro2b) [process, right of=dec1, xshift=3.5cm] {Swap atomic positions};
\node (dec2) [decision, below of=pro2b,yshift=-1.5cm] {$\mathrm{\Vert B_{ideal} - B_{real} \Vert} = 0$};
\draw [arrow] (start) -- (in1);
\draw [arrow] (in1) -- (pro1);
\draw [arrow] (pro1) -- (dec1);
\draw [arrow] (dec1) -- node[anchor=east]{Yes}(pro2a);
\draw [arrow] (dec1) -- node[anchor=south] {No} (pro2b);
\draw [arrow] (pro2b) -- (dec2);
\draw [arrow] (dec2) -- node[anchor=south]{Yes}(pro2a);
\draw [arrow] (dec2) -- node[above,near start]{No} ++(3.4,0) |- (pro2b);

\end{tikzpicture}
}
\caption{Algorithm for the random structure generation}
\label{fig:random}
\end{figure}

\begin{figure*}
    \centering
    \begin{subfigure}{\textwidth}
        \includegraphics[width=\textwidth]{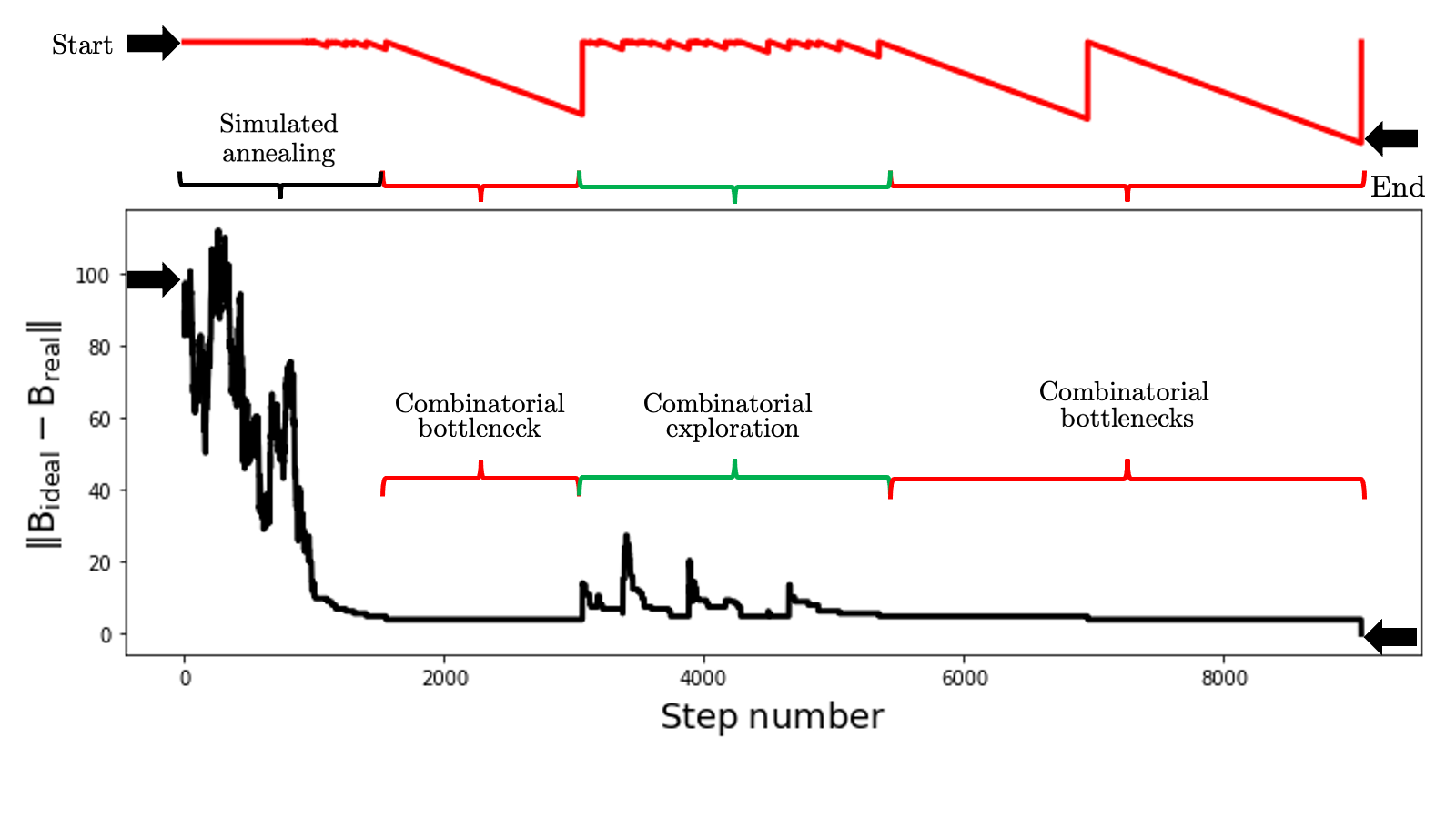}
        \caption{}
        \label{fig:schematic}
    \end{subfigure}
    \begin{subfigure}{0.45\textwidth}
        \includegraphics[width=\textwidth]{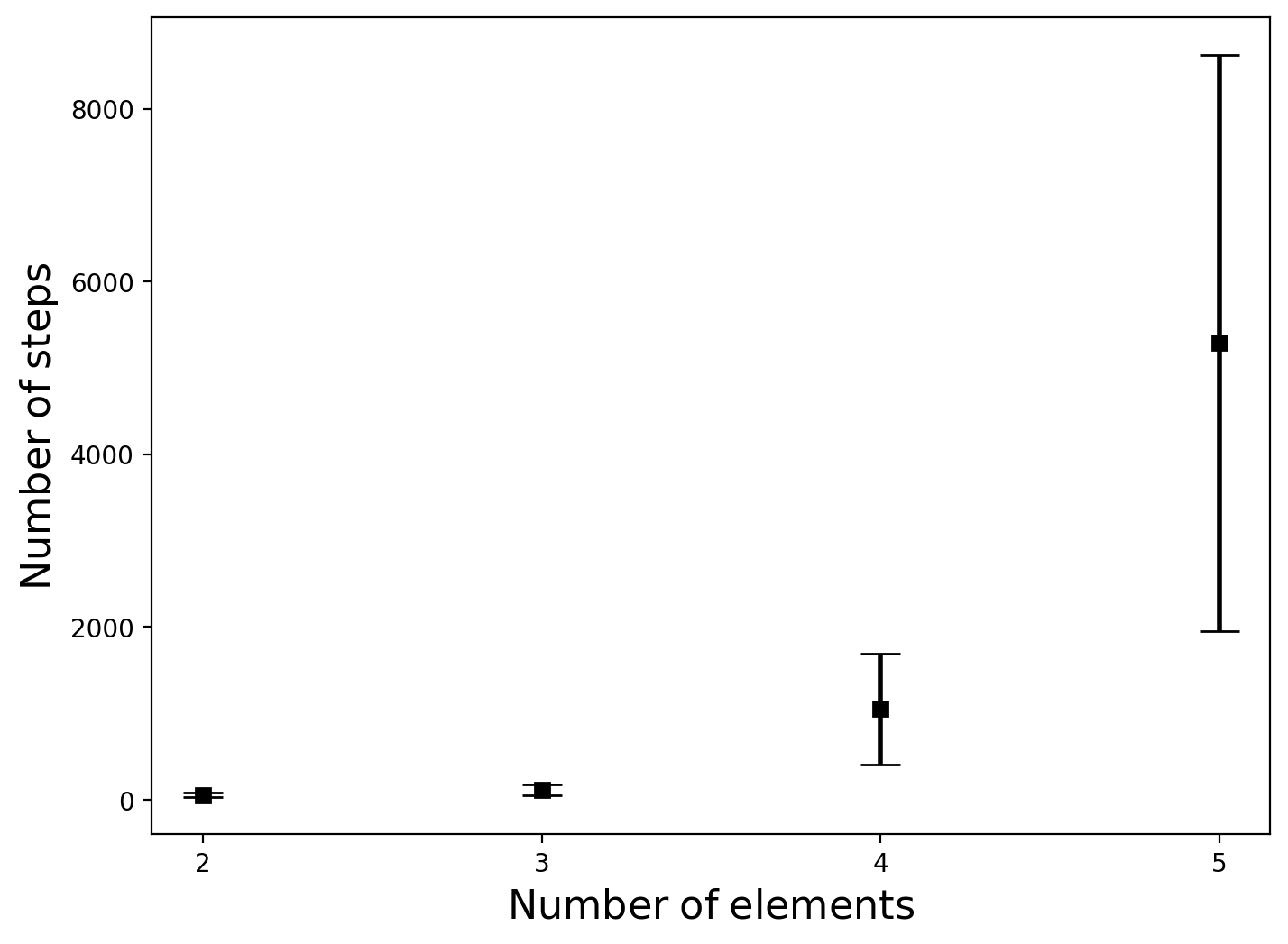}
        \caption{}
        \label{fig:elem_num_steps}
    \end{subfigure}
    \begin{subfigure}{0.45\textwidth}
        \includegraphics[width=\textwidth]{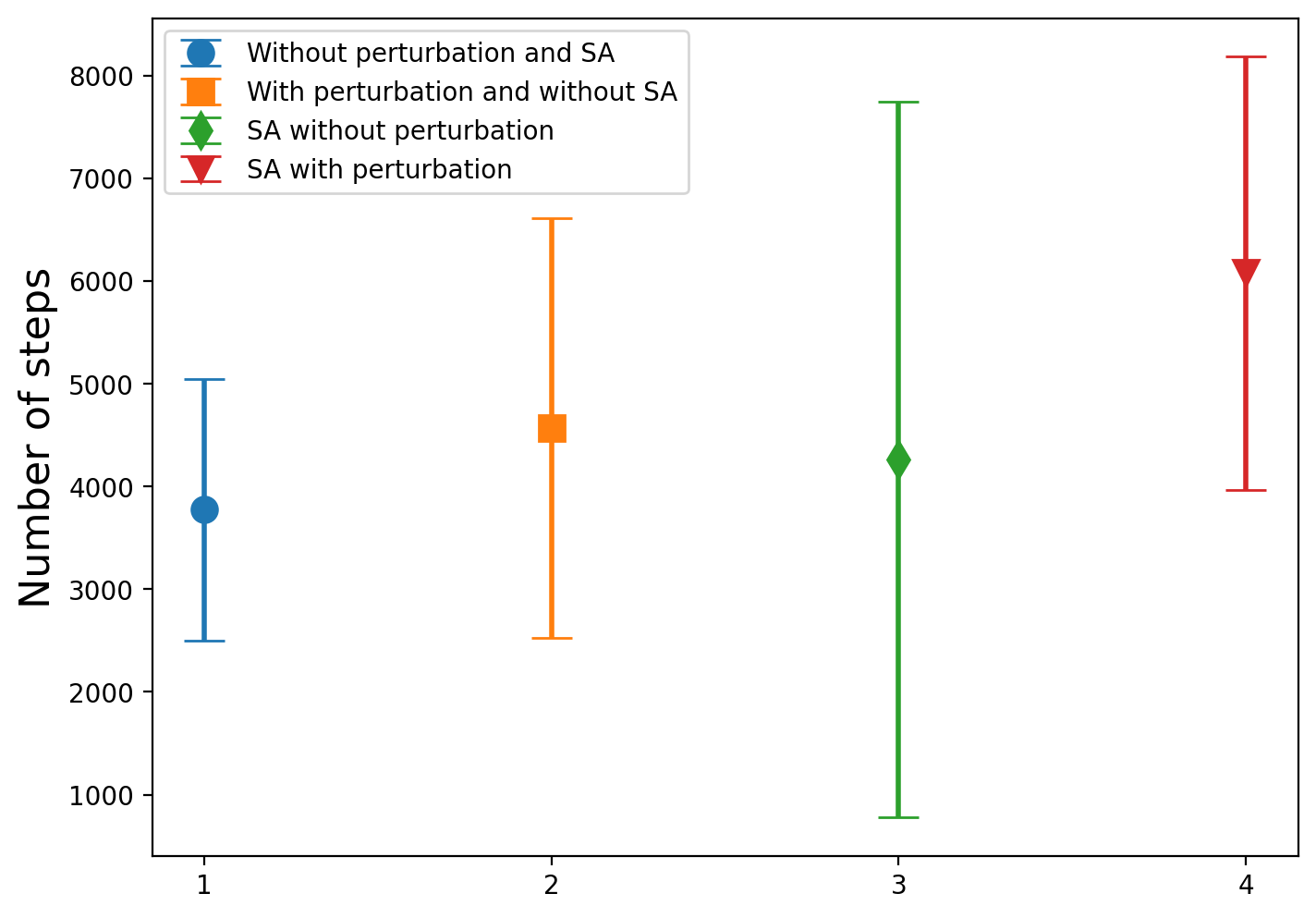}
        \caption{}
        \label{fig:sa_effect}
    \end{subfigure}
    \caption{\ref{fig:schematic} Schematic showing the variation in $\mathrm{\Vert B_{ideal} - B_{real} \Vert}$ and corresponding change in the bottleneck parameter (as shown schematically on top of the figure) for combined simulated annealing and combinatorial exploration for generation of random structure. Note that the increasing depth of the top line represents the combinatorial bottlenecks, \ref{fig:elem_num_steps} shows the number of steps required to generate the perfectly chemically disordered or random structures for 600 atoms supercell. Five structures for each case was generated. \ref{fig:sa_effect} shows the effect of the simulated annealing (SA) and combinatorial perturbation for 600 atoms supercell. 1, 2, 3 and 4 in x-axis represents generation of the random structure of 5-component alloy without SA and perturbation, with perturbation and without SA, SA without perturbation and SA with perturbation, respectively. 10 configurations for each of these cases were generated.}
    \label{fig:SA_CB}
\end{figure*}

The generation of random structure is the first step towards creating a structure with predefined short-range order. The generation of the chemically disordered or random structure, as shown in Fig. \ref{fig:random} is carried out in the sequential manner.
The first step of the random structure generation involves the random shuffling of atomic positions. It generates the structure with random atoms sitting at the crystal lattice sites, generating a pseudo-random structure. The generated structure is quantified in terms of the vector, $\mathrm{B_{real}}$, which may be represented as:
    \begin{equation}
        \mathrm{
        B_{real} = \left[ m_{ii}\;m_{ij}\; \cdot\; \cdot\; \cdot\; \cdot\; m_{kk} \right]
        }
    \end{equation}
where, $\mathrm{m_{ii}}$ represents the like-bond number or bond between similar type of chemical species ($\mathrm{\emph{i.e.,}\; i}$), while $\mathrm{m_{ij}}$ is unlike-bond number or chemical bond between unlike chemical species ($\mathrm{\emph{i.e.,}\; i\; \& \; j}$, where $\mathrm{i \neq j}$). The $\mathrm{m_{ii}}$ or $\mathrm{m_{ij}}$ in the present investigation have been determined using the \emph{neighbourlist} functionality of the Atomistic Simulation Environment (ASE) \cite{Hjorth_Larsen_2017}. 
The difference of the pseudo-random configuration generated through random shuffling is quantified in terms of the euclidean distance between the $\mathrm{B_{real}}$ and $\mathrm{B_{ideal}}$. The $\mathrm{B_{ideal}}$ is the vector containing the ideal bond numbers for like and unlike bonds for the complete chemical disorder. The ideal bond number ($\mathrm{n}$) for like-bond is defined by equation \ref{two_eq}, while the ideal bond-number for the unlike-bond is $\mathrm{2\cdot n}$.
 The pseudo-random structure generated through random shuffling is further randomised through the swapping of atomic positions to ensure that the swapping leads to the lower value of $\mathrm{\Vert B_{ideal} - B_{real} \Vert}$. The simulated annealing (SA) of initially pseudo-random structure is initially carried out, where configurations are heated to $\mathrm{0.1 \cdot \Vert B_{ideal} - B_{real} \Vert}$  and then these configurations are cooled to $\mathrm{0}$ through linear cooling protocol in $\mathrm{10 \cdot \Vert B_{ideal} - B_{real} \Vert}$ steps  (Figure \ref{fig:random}). Such swapping sequentially leads towards the random structure. However, if the structure is stuck in the \emph{combinatorial bottleneck} (\emph{i.e.,}, when any swapping does not lead to $\mathrm{\Vert B_{ideal} - B_{real} \Vert} = 0$), then exploration of other combinatorial landscape is allowed (perturbation).  Once the configuration has reached the condition of $\mathrm{\Vert B_{ideal} - B_{real} \Vert} = 0$, the swapping process is stopped (Fig. (\ref{fig:schematic})). It can be seen in Figure (\ref{fig:elem_num_steps}) that number of steps for attaining the perfectly random structure increases with the increase in the number of elements in an alloy. It is understandable, as the number of elements in the alloy increase the combinatorial complexity increases. Fig. (\ref{fig:sa_effect}) shows the influence of SA and perturbation on the number of steps required for generating the chemical disordered configuration for a 5-component alloy. It should be noted that even though SA and perturbation leads to the increased number of steps for generation of perfectly chemically-disordered structure, but it ensures that a system is not stuck in the combinatorial bottleneck indefinitely.  

\section*{Generation of configurations with desired CSRO}
\begin{figure*}
    \centering
    \begin{subfigure}{0.8\textwidth}
        \includegraphics[width=\textwidth]{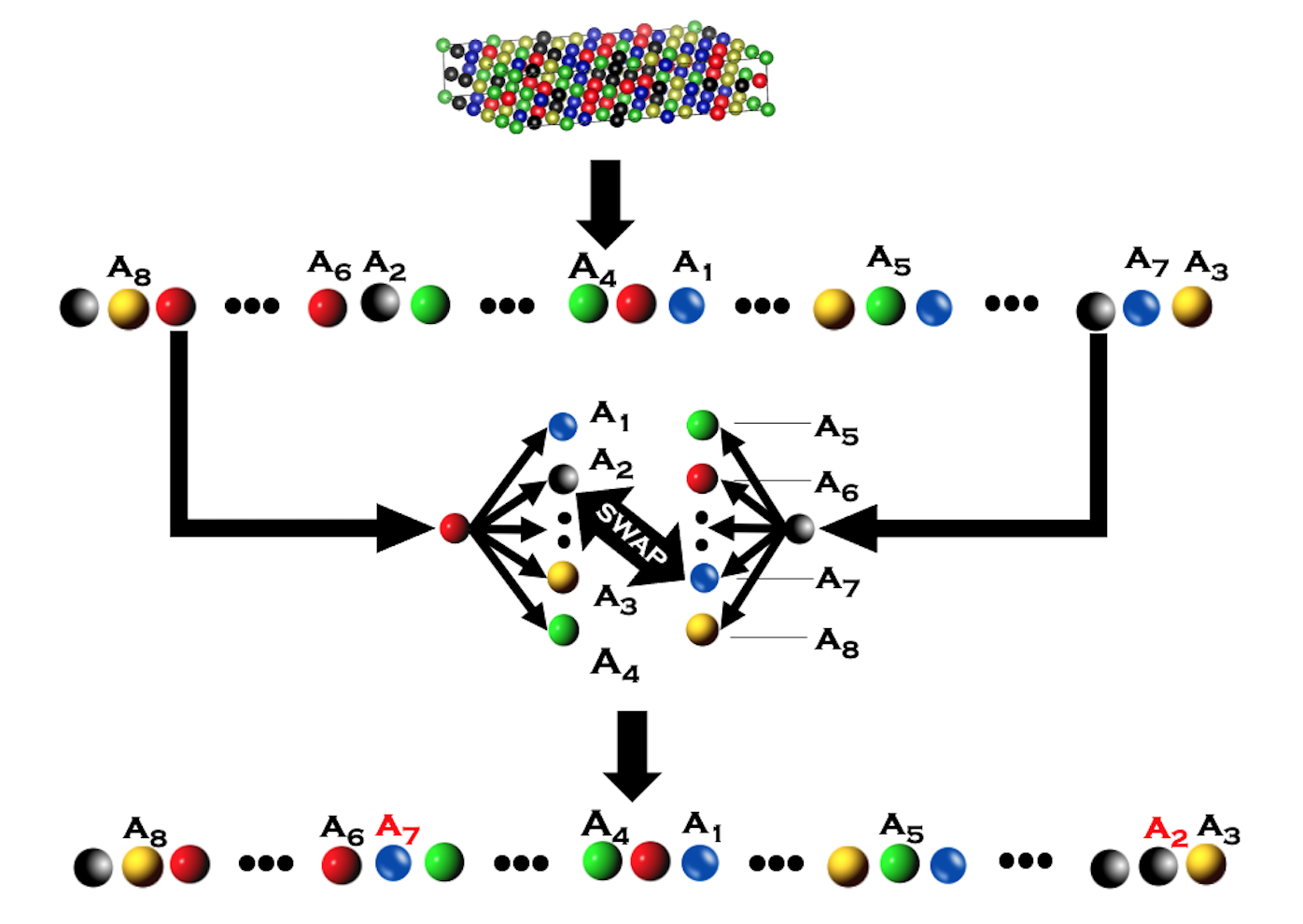}
        \caption{}
        \label{fig:sro_gen}
    \end{subfigure}
    \begin{subfigure}{\textwidth}
        \resizebox{\textwidth}{!}{
        \begin{tikzpicture}[node distance=2cm]
        \centering
        \node (start) [startstop] {AC, BD $\mathrm{\xrightarrow{swap}}$ AD, BC};
        \node (pro1)[process,below of=start]{A is $\mathrm{CA_1}$, B is $\mathrm{CA_2}$, C is $\mathrm{SP_1}$ and D is $\mathrm{SP_2}$}; 
        \node (pro2)[process, below of= pro1,xshift=-9.5cm]{Find $\mathrm{CA_1}$ atom}; %
        \node (pro3) [process,below of = pro2]{Find  $\mathrm{CN_1}$};
        \node (dec1) [decision,below of = pro3]{$\mathrm{SP_1}$ in $\mathrm{CN_1}$};
        \node (pro4) [process, below of = pro1,xshift=-5.5cm]{Find $\mathrm{CA_2}$ atom}; %
        \node (pro5) [process, below of = pro4]{Find  $\mathrm{CN_2}$};
        \node (dec2) [decision,below of = pro5]{$\mathrm{SP_2}$ in $\mathrm{CN_2}$};
        \node (end1) [startstop,below of = dec1]{X};
        \node (pro6) [process,below of = dec2]{Find NN of $\mathrm{SP_2}$ $\mathrm{\left(NN_1 \right)}$};
        \node (dec3) [decision,below of = pro6]{$\mathrm{CA_1}$ in $\mathrm{NN_1}$};
        \node (end2) [startstop,below of = dec3]{Y};

        \draw[arrow] (start) -- (pro1);
        \draw [arrow] (pro1) -| node[below,near end,anchor=west] {\textbf{AD is the preferred-bond}} (pro2);
        \draw [arrow] (pro1) -- (pro4);
        \draw [arrow] (pro2) -- (pro3);
        \draw [arrow] (pro3) -- (dec1);
        \draw [arrow] (pro4) -- (pro5);
        \draw [arrow] (pro5) -- (dec2);
        \draw [arrow] (dec1) -- node[anchor=east]{Yes}(end1);
        \draw [arrow] (dec2) -- node[anchor=east]{Yes}(pro6);
        \draw [arrow] (pro6) -- (dec3);
        \draw [arrow] (dec3) -- node[anchor=east]{No}(end2);

        \node (pro7)[process, below of= pro1,xshift=5.5cm]{Find $\mathrm{CA_1}$ atom}; %
        \node (pro8) [process,below of = pro7]{Find  $\mathrm{CN_1}$};
        \node (dec4) [decision,below of = pro8]{$\mathrm{SP_1}$ in $\mathrm{CN_1}$};
        \node (pro9) [process, below of = pro1,xshift=9.5cm]{Find $\mathrm{CA_2}$ atom}; %
        \node (pro10) [process, below of = pro9]{Find  $\mathrm{CN_2}$};
        \node (dec5) [decision,below of = pro10]{$\mathrm{SP_2}$ in $\mathrm{CN_2}$};
        \node (end4) [startstop,below of = dec5]{Y};
        \node (pro11) [process,below of = dec4]{Find NN of $\mathrm{SP_1}$ $\mathrm{\left(NN_1 \right)}$};
        \node (dec6) [decision,below of = pro11]{$\mathrm{CA_2}$ in $\mathrm{NN_1}$};
        \node (end3) [startstop,below of = dec6]{X};

        \draw [arrow] (pro1) -| node[below,near end,anchor=east] {\textbf{BC is the preferred-bond}}(pro9);
        \draw [arrow] (pro1) -- (pro7);
        \draw [arrow] (pro7) -- (pro8);
        \draw [arrow] (pro8) -- (dec4);
        \draw [arrow] (dec4) -- node[anchor=west]{Yes}(pro11);
        \draw [arrow] (pro11) -- (dec6);
        \draw [arrow] (dec6) -- node[anchor=west]{No}(end3);
        \draw [arrow] (pro9) -- (pro10);
        \draw [arrow] (pro10) -- (dec5);
        \draw [arrow] (dec5) -- node[anchor=west]{Yes} (end4);
    \end{tikzpicture}
    }
        \caption{}
        \label{fig:swap_schematic}
    \end{subfigure}
    \caption{Algorithm for generation of chemical short range order from the random structure. \ref{fig:sro_gen} showing the schematic of the CSRO generation algorithm and \ref{fig:swap_schematic} showing the details of the swap process. Note that $\mathrm{CN_1}$  and $\mathrm{CN_2}$ are nearest-neighbours of $\mathrm{CA_2}$  and $\mathrm{CA_2}$, respectively. Also, NN in the figure is the abbreviation of the nearest-neighbour.}  
\end{figure*}

The CSRO is imposed on the perfectly chemically-disordered supercell with the compositional constraint. To keep the composition constant, swap algorithm as shown in Figure (\ref{fig:sro_gen} and \ref{fig:swap_schematic}) has been devised. Initially, dependent upon the preferred-bond (whose propensity needs to be increased), two groups of bonds are determined (group-1 and group-2). For example, in a ABCD alloy, if the A-D bond is the preferred-bond, then swap between (A-A, C-D), (A-B, D-D) as well as (A-C, B-D) can increase the preferred-bond propensity. In this case, A-A, A-B and A-C can belong to the group-1, while C-D, D-D and B-D bonds constitute the group-2. One atom from the group-1 bond, which forms the preferred bond and another atom from group-2 which is not the part of the preferred bond forms the \emph{central pair} or $\mathrm{[CA_1,CA_2]}$, respectively. While, left atom in group-1 bond and group-2 bonds form \emph{swap pair} or $\mathrm{[SP_1,SP_2]}$, respectively. Next the nearest-neighbour of $\mathrm{CA_1}$ ($\mathrm{CN_1}$) and $\mathrm{CA_2}$ ($\mathrm{CN_2}$) are determined. It should be noted that in each of the swap iteration, newly changed atomic positions are interred into the \emph{forbidden-list} or $\mathrm{FL}$. The idea behind this list is to put the constraint that once a atomic swap has been carried out to generate the required preferred-bond, it should not be changed in the subsequent iterations. Suppose the swap between bonds AC and BD yields AD and BC or $\mathrm{swap(AC,BD)} \rightarrow \mathrm{AD,BC}$. Depending upon whether AD or BC is the preferred bond, slightly different swap algorithm is employed.\\
In the $\mathrm{CN_1}$, there is a possibility that some of the atoms might already be in the $\mathrm{FL}$, while  other atoms might include the $\mathrm{SP_1}$ and $\mathrm{SP_2}$. If the preferred-bond is AD, then the $\mathrm{CA_1}$ is A. $\mathrm{SP_2}$ is directly interred into FL, as $\mathrm{CA_1}$ and $\mathrm{SP_2}$ already are forming the preferred-bond or AD. $\mathrm{SP_1}$ atoms in the $\mathrm{CN_1}$ are swapable atoms in $\mathrm{CN_1}$ ($\mathrm{X}$). Similarly, $\mathrm{CA_2}$ is B, whose nearest-neighbours are $\mathrm{CN_2}$. $\mathrm{SP_2}$ atoms in $\mathrm{CN_2}$ are determined, and if any of these atoms have $\mathrm{CA_1}$ as their nearest-neighbour, those are interred into FL, while other $\mathrm{SP_2}$ atoms constitute the swapable atoms in $\mathrm{CN_2}$ ($\mathrm{Y}$). The number of swaps possible are then $\mathrm{\vert X - Y\vert}$.\\
If, BC is the preferred-bond in the above-stated swap scenario, in that case, $\mathrm{SP_1}$ atoms in $\mathrm{CN_1}$ are checked, if they have $\mathrm{CA_2}$ as their nearest-neighbours. Such $\mathrm{SP_1}$ atoms are added to the FL, while the remaining $\mathrm{SP_1}$ atoms are swapable atoms in the $\mathrm{CN_1}$. In $\mathrm{CN_2}$, $\mathrm{SP_1}$ atoms are added to FL, while $\mathrm{SP_1}$ are swapable atoms in $\mathrm{CN_2}$ or $\mathrm{Y}$. The number of swaps possible in such scenario is again $\mathrm{\vert X - Y\vert}$. \\

\begin{figure*}
    \centering
    \includegraphics[width=0.8\textwidth]{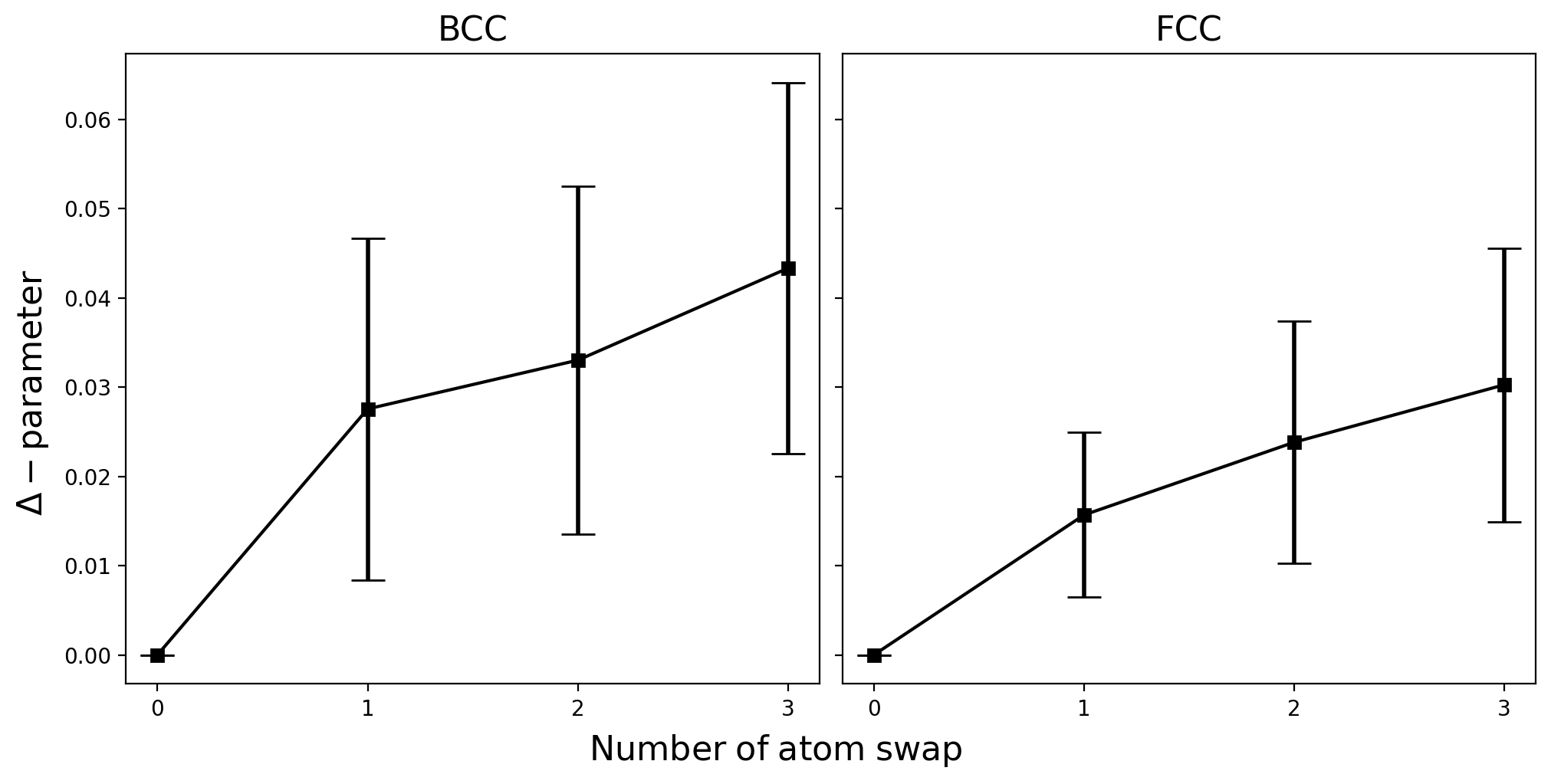}
    \caption{Testing $\mathrm{\Delta-parameter}$ for Landau's first condition for the order parameter}
    \label{fig:landau}
\end{figure*}
\begin{figure*}
    \centering
    \begin{subfigure}{\textwidth}
        \centering
        \includegraphics[width=\textwidth]{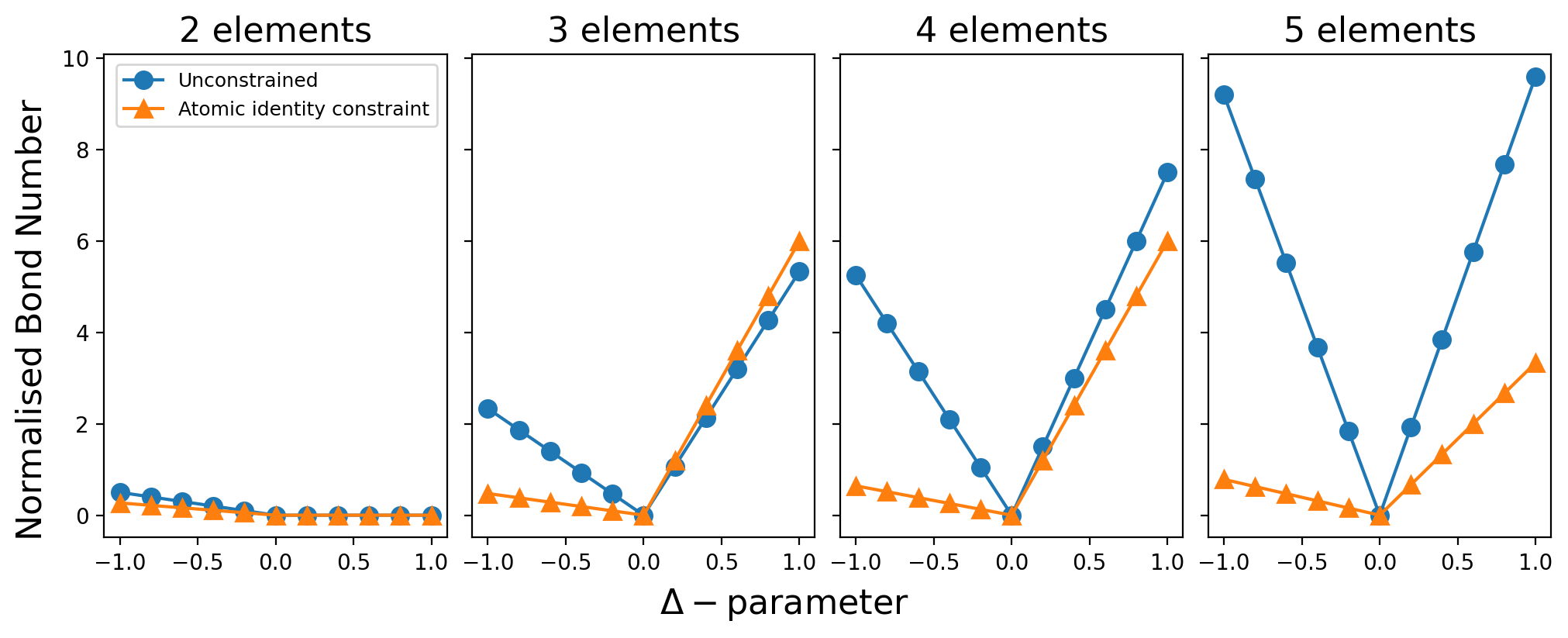}
        \caption{}
        \label{fig:norm_bond}
    \end{subfigure}
    \begin{subfigure}{\textwidth}
        \centering
        \includegraphics[width=\textwidth]{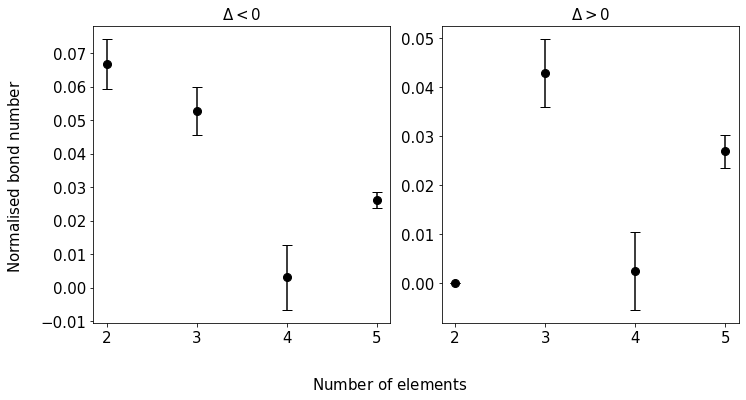}
        \caption{}
        \label{fig:bond_ele}
    \end{subfigure}
    
    \caption{\ref{fig:norm_bond} Variation of the number of preferred bonds (like and unlike, depending upon the $\mathrm{\Delta}$)) for systems containing different number of elements. The y-axis represents the normalised bond number, which is the value of number of bonds which need to be increased in addition to the number of bonds at the perfect chemical disorder, \emph{i.e.,} $\mathrm{n}$ for like-bonds and $\mathrm{2n}$ for unlike-bonds, divided by the either $\mathrm{n}$ (for like-bonds) or $\mathrm{2n}$ (for unlike-bonds) and \ref{fig:bond_ele} normalised bond number for different number of elements, as full compositional constraint is applied in the OPERA framework.} 
    \label{fig:bond_vals}
\end{figure*}

Figure (\ref{fig:landau}) shows the variation of $\mathrm{\Delta-parameter}$ for body-centered cubic (BCC) and face-centered cubic (FCC) lattice for 180 atom supercell. 1000 runs were carried out sequentially for 1 to 3 atom ordering. It can be seen that statistically $\mathrm{\Delta-parameter}$ is different for BCC and FCC phase, which is the first condition as depicted in Landau's theory \cite{callen1998thermodynamics}. We additionally demonstrate the effect of various constraints (\emph{i.e.,} no constraint, atomic identity constraint and full constraints involving the \emph{forbidden-lists}) on the number of swaps possible, while keeping the composition constant. \\
Initially, we analytically study the number of atoms which may be swapped, if the identity of the bonds is simply like-bond and unlike bond, without any consideration to the identity of atoms forming these bonds. As mentioned in the section-2 of SI, the change in the preferred bond number required for $\mathrm{\Delta > 0}$ (the preferred bond is a like-bond) is,
\begin{equation}
    \mathrm{
    \centering
    a = \frac{2 \cdot K_2 \cdot \Delta \cdot n \cdot \left(2K_1 + K_2 - 1 \right)}{\left(2 \cdot 
    K_1 + K_2  \right)}
    }
\end{equation}
While for $\mathrm{\Delta < 0}$, the required change in the preferred unlike-bonds is,
\begin{equation}
    \mathrm{
    \centering
    a = -\frac{2 \cdot K_1 \cdot \Delta \cdot n \cdot \left(2K_1 + K_2 - 1  \right)}{\left(2 \cdot 
    K_1 + K_2  \right)}
    }
\end{equation}
While, if the bond-identity is introduced, the concept of the preferred-bond becomes more specific in the comparison of the above-stated case, where only like and unlike bonds were considered. The identity of the preferred-bond further narrows down the bonds which may be swapped to generate the preferred bond. As shown in the section-3 of SI, the number of bond-swaps for $\mathrm{\Delta < 0}$ may be given as,
\begin{equation}
   \mathrm{
   \centering
   a = \frac{\Delta \cdot K_1 \cdot K_2 \cdot n \cdot \left(-5e -4\right)}{\left(e^2 \cdot K_1 + 7 e^2 \cdot K_2 - 4 e \cdot K_1\right)}
   } 
\end{equation}
If $\mathrm{\Delta > 0}$, then,
\begin{equation}
    \mathrm{
    \centering
    a = \frac{2 \cdot \Delta \cdot K_1 \cdot K_2 \cdot n}{\left(2 \cdot K_1 +  K_2\right) \cdot \left(e - \floor{\frac{e}{2}} - 1 \right)}
    }
\end{equation}
where, $\mathrm{e}$ is the number of type of atoms in the system. Note that $\floor{}$ in the above expression represents the floor operator. It can be seen in Fig. (\ref{fig:norm_bond}) that as the number of elements in the \emph{unconstrained case} increases, the required number of like as well as unlike bonds increases, which implies that as the number of the type of elements in the material increases, it becomes increasingly harder to attain the complete ordering in both sides of the disorder. As the atomic identity constraint is introduced and in addition to the identity of bond (like or unlike), the atomic identity of bond forming elements becomes crucial, the number of bonds, which may be swapped for desired ordering is also reduced (beyond 3 elements). Note that in case of binary system, $\mathrm{\Delta > 0}$ is not possible is the composition of the supercell needs to be constant, as the only bonds available except like bonds is an unlike bond. The swap between this bond cannot lead to the desired like bond. So, normalised bond number is zero for $\mathrm{\Delta > 0}$.  For the ternary system, it should be noted that for $\mathrm{\Delta < 0}$, the normalised bond number for the atomic identity constraint is  lower than that of unconstrained case. However, for $\mathrm{\Delta > 0}$, it can be seen that normalised bond number for the unconstrained case is lower than the atomic identity constraint. Such observation can be rationalised in terms of value of unlike-bonds $\mathrm{(K_1)}$ and like-bonds $\mathrm{(K_2)}$. Ternary systems are the only case, where $\mathrm{K_1 = K_2}$. Figure \ref{fig:bond_ele} shows the normalised bond number for binary, ternary, quaternary and quinary systems as determined by carrying out 1000 runs for each case to determine the maximum bond change possible using OPERA code. However, both of these analytical schemes provide the upper bounds for the number of bonds desired to attain the particular $\mathrm{\Delta}$ value. But in reality as we introduce the constraints in the OPERA scheme, where the composition of the supercell need to be constant, the bond numbers which can be altered is 2 order of magnitude lower than the upper bounds as discussed earlier. But it should be noted that normalised bond number is not the monotonic function of the number of type of elements in the system. Rather, it is determined by the combinatoric landscape of the multicomponent system. (Fig. \ref{fig:bond_ele}).     

\section*{Comparison of the $\mathrm{\Delta}$-parameter with the Warren-Cowley CSRO parameter ($\mathrm{\alpha_{ij}}$)}
\begin{figure*}
    \centering
    \includegraphics[width=\textwidth]{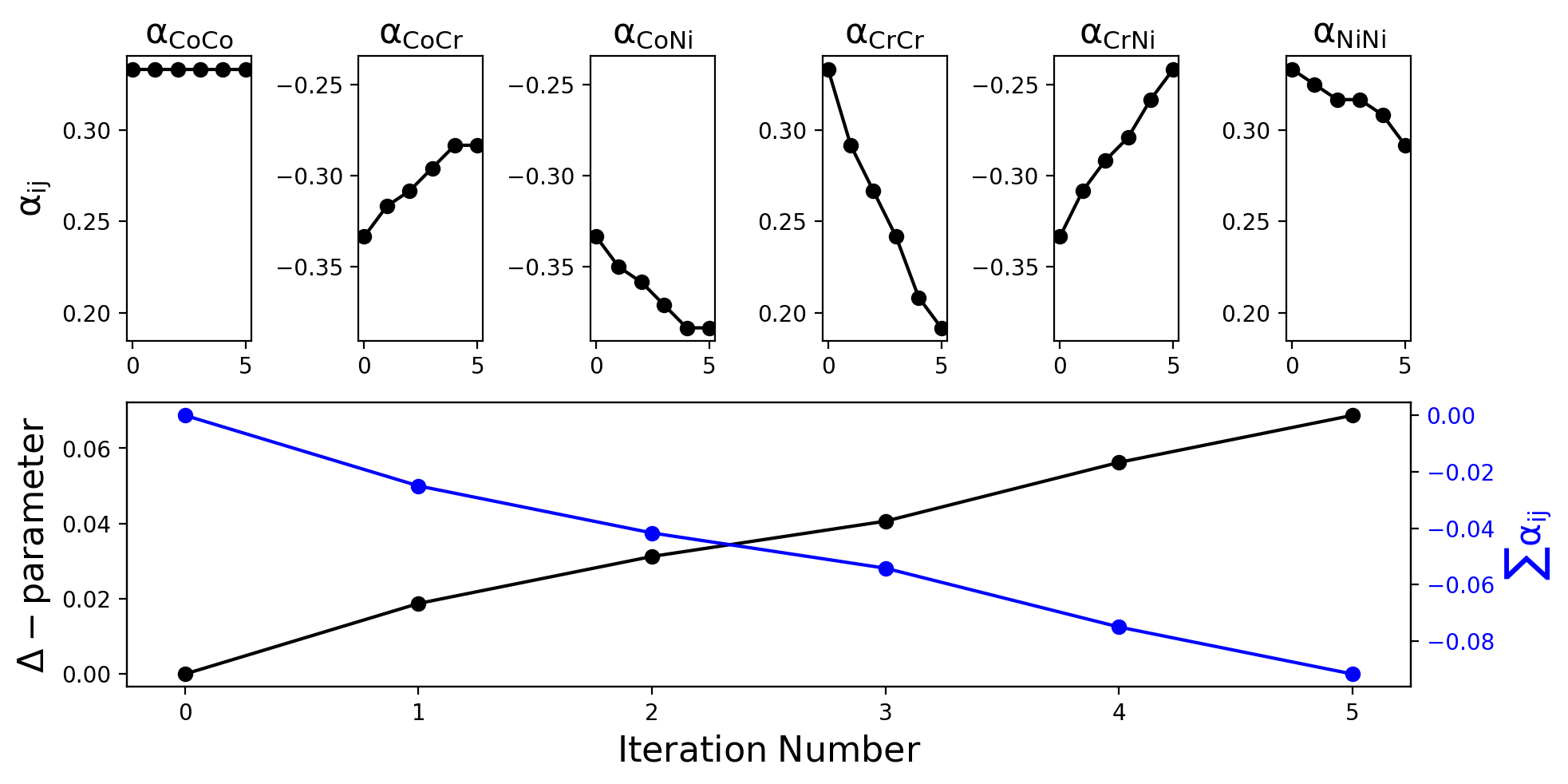}
    \caption{Comparison of Warren-Cowley order parameter ($\mathrm{\alpha_{ij}}$) calculated for a configuration of FCC-CoCrNi alloy with $\mathrm{\Delta}$-parameter calculated for the same configuration. Note that x-axis of each plot correspond to the iteration number, which corresponds to the increased number of Cr-Cr bonds sequentially. In this case, with increase in Cr-Cr bond number, Co-Cr and Cr-Ni bond numbers decrease and Co-Ni and Ni-Ni bonds increase, while the Co-Co bond number remains constant during the iteration.}
    \label{fig:wc_delta}
\end{figure*}

Figure \ref{fig:wc_delta} shows the variation of $\mathrm{\alpha_{ij}}$ parameters for a configuration of a CoCrNi alloy, where the propensity of Cr-Cr bonds were sequentially increased from the chemically disordered value (\emph{i.e.,} $\mathrm{n}$ as given by equation (\ref{two_eq})) to the maximum value of Cr-Cr bonds possible in the 5-steps. Since, we would like to maintain the composition of the supercell to generate desired CSRO, there are associated changes in the bond-numbers. In the above-stated case of CoCrNi, with the increase in Cr-Cr bond, there is decrease in the Co-Cr and Cr-Ni bonds, while there is increase in Co-Ni and Ni-Ni bonds. The Co-Co bond numbers remain constant. For such a scenrio, It can be seen that $\mathrm{\Delta}$-parameter increases linearly with increase in the number of Cr-Cr bonds. However, it should be noted that even the Cr-Cr bonds are increasing, the value of $\mathrm{\alpha_{CrCr}}$ is decreasing, with is counter-intuitive. However, as it can be seen in equation (\ref{eq:wc}) that the Warren-Cowley CSRO parameter calculation for like-bond would involve the $\mathrm{1-\dfrac{P_{i|i}}{c_i}}$ expression, where the $\mathrm{P_{i|i}}$ is the probability of finding atom $\mathrm{i}$ in the required coordination shell of $\mathrm{i}$, while $\mathrm{c_i}$ is the concentration of the same atom. Note that, in the present discussion, we are keeping the concentration of the ensemble of atom constant and due to the even though with the increase in Cr-Cr bonds, the $\mathrm{P_{i|i}}$ term is increasing, but the $\mathrm{c_i}$ is constant. So, the value of $\mathrm{\alpha_{CrCr}}$ bond is decreasing, even the Cr-Cr bond propensity is increasing. Such observation points towards the fundamental limitation of Warren-Cowley CSRO for dealing with constant composition configurations, as it requires the semi-grand canonical ensemble in this scenario. But, it can be seen that $\mathrm{\Delta}$-parameter is monotonically increasing with the increase in the Cr-Cr bond propensity in canonical ensemble (while the composition is constant) and $\mathrm{\sum \alpha_{ij}}$ is decreasing monotonically. Hence, the $\Delta$ parameter can quantify the like-ordering in the above-stated scenario.        
\section*{Validation of proposed $\mathrm{\Delta}$ parameter}
\begin{figure*}
    \centering
    \begin{subfigure}{0.4\textwidth}
        \centering
        \includegraphics[width=\textwidth]{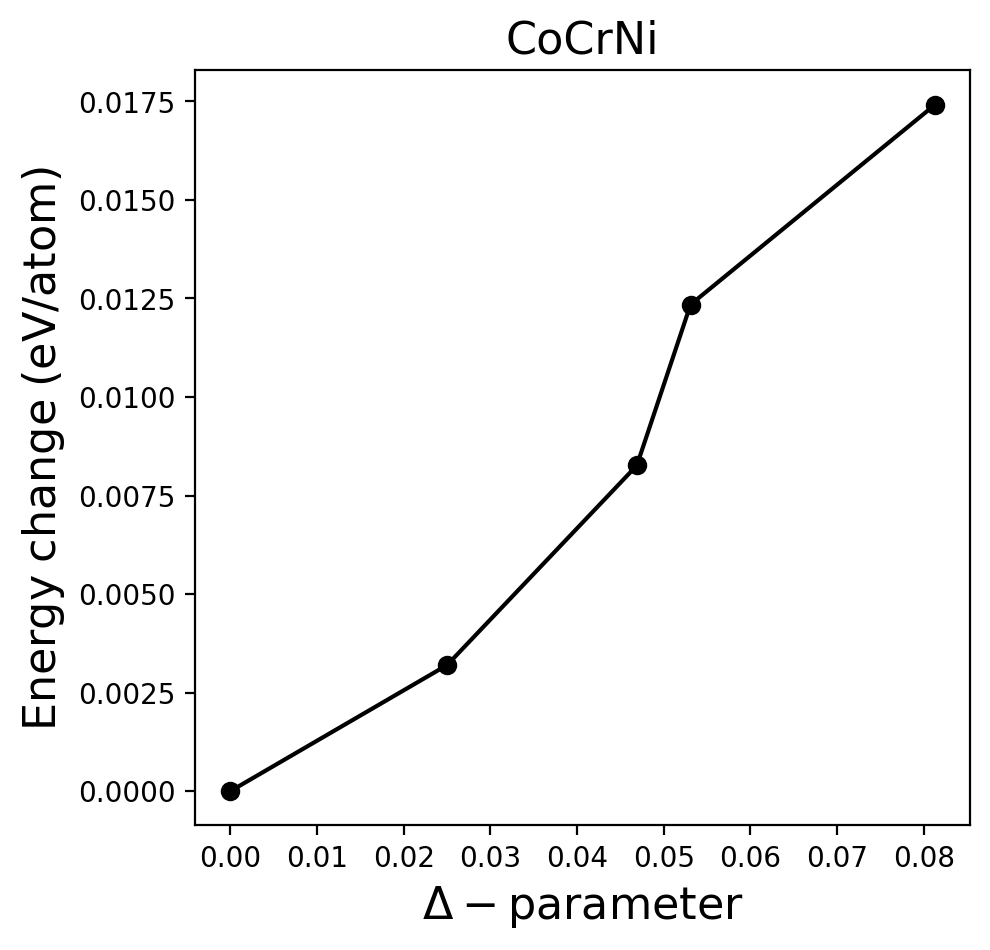}
        \caption{}
        \label{fig:cr_cr}
    \end{subfigure}
        \begin{subfigure}{0.4\textwidth}
        \centering
        \includegraphics[width=\textwidth]{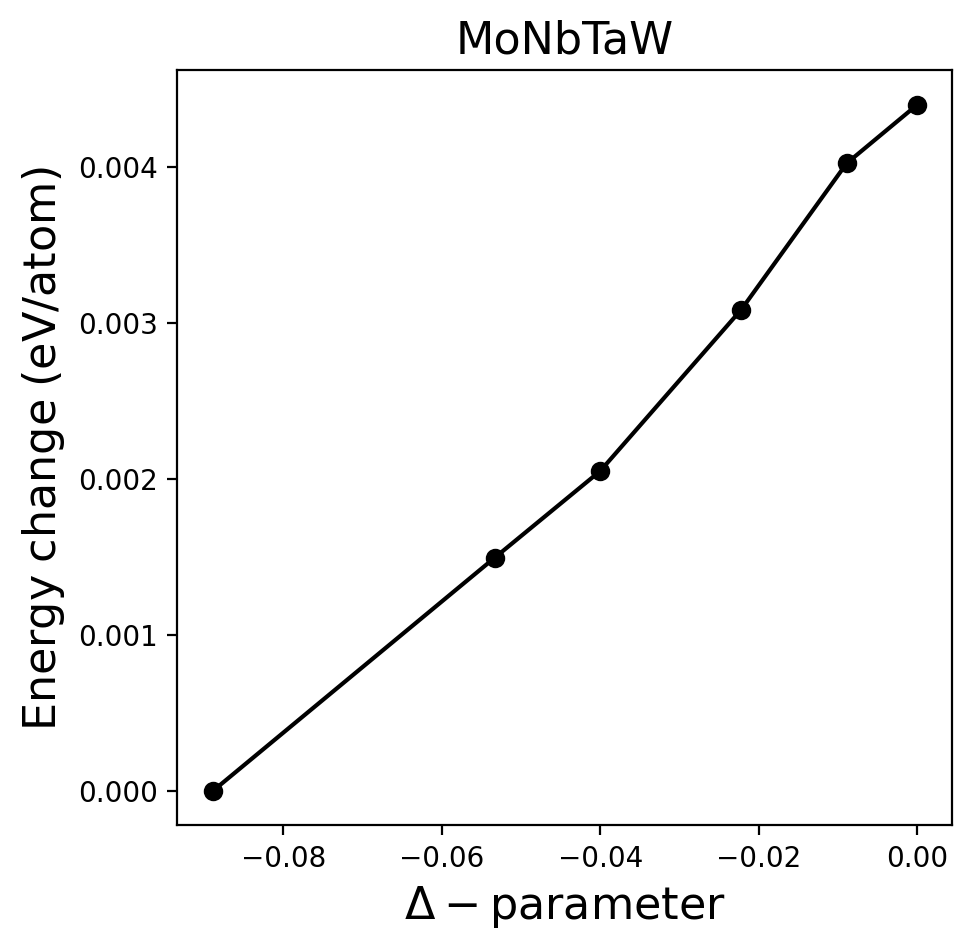}
        \caption{}
        \label{fig:mo_ta}
    \end{subfigure}
    \caption{(\ref{fig:cr_cr}) Variation in energy with increase in $\Delta$ due to greater Cr-Cr bonds and (\ref{fig:mo_ta}) Decrease in the energy of MoNbTaW HEA due to increase in Mo-Ta bonds, leading to decrease in $\mathrm{\Delta}$ parameter.}
\end{figure*}

\begin{figure*}
    \centering
    \begin{subfigure}[b]{0.3\textwidth}
        \centering
        \includegraphics[width=\textwidth]{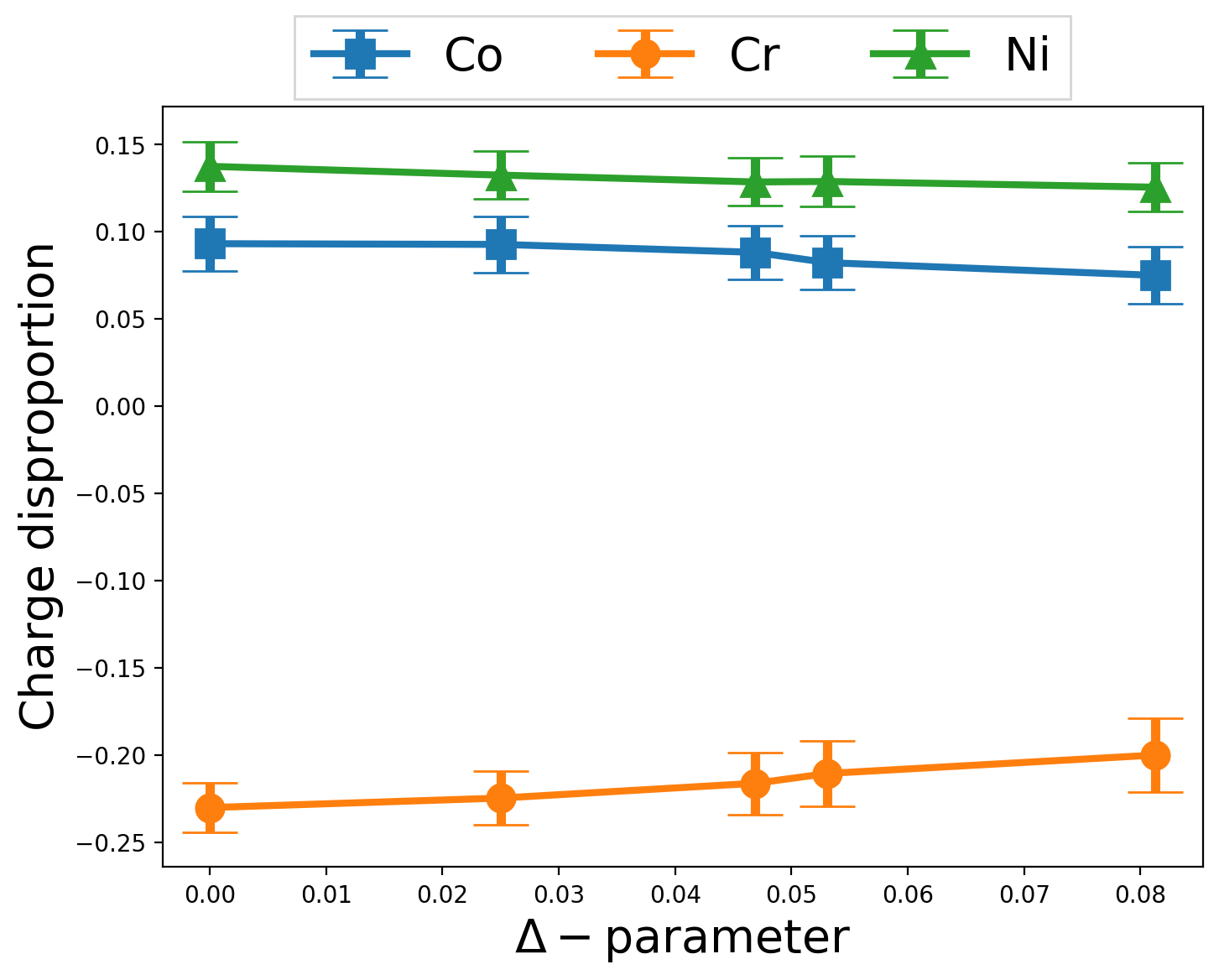}
        \caption{}
        \label{fig:ccn_charge}
    \end{subfigure}
    \begin{subfigure}[b]{0.3\textwidth}
        \centering
        \includegraphics[width=\textwidth]{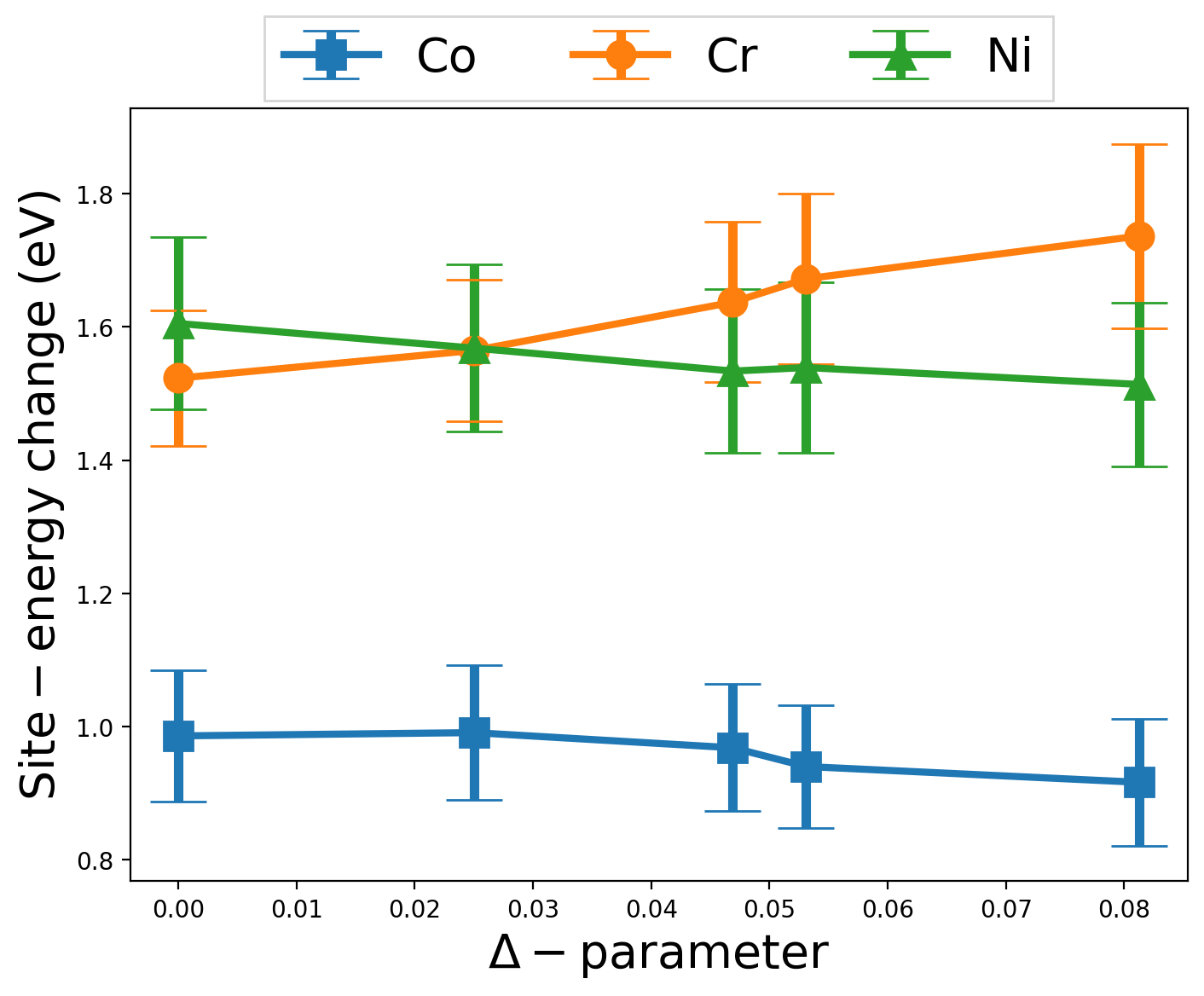}
        \caption{}
        \label{fig:ccn_siteene}
    \end{subfigure}
    \begin{subfigure}[b]{0.3\textwidth}
        \centering
        \includegraphics[width=\textwidth]{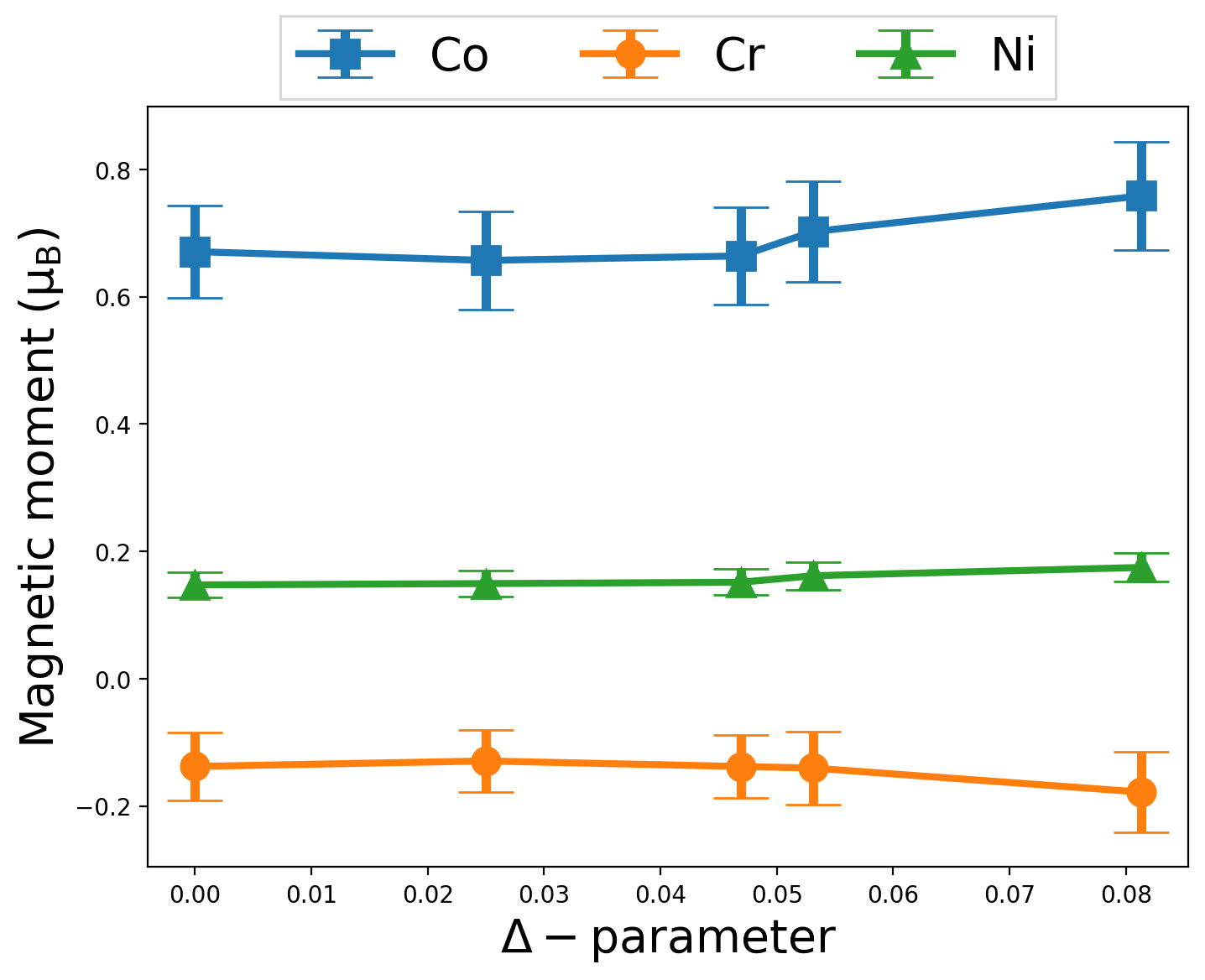}
        \caption{}
        \label{fig:ccn_magmom}
    \end{subfigure}
    \caption{The variation in the (\ref{fig:ccn_charge}) charge disproportion, (\ref{fig:ccn_siteene}) atomic site-energies and (\ref{fig:ccn_magmom}) atomic magnetic moments for CoCrNi alloy. Note that error bars represents 95\% confidence interval here.}
    \label{fig:ccn_analysis}
\end{figure*}

\begin{figure*}
    \centering
    \begin{subfigure}[b]{0.3\textwidth}
        \centering
        \includegraphics[width=\textwidth]{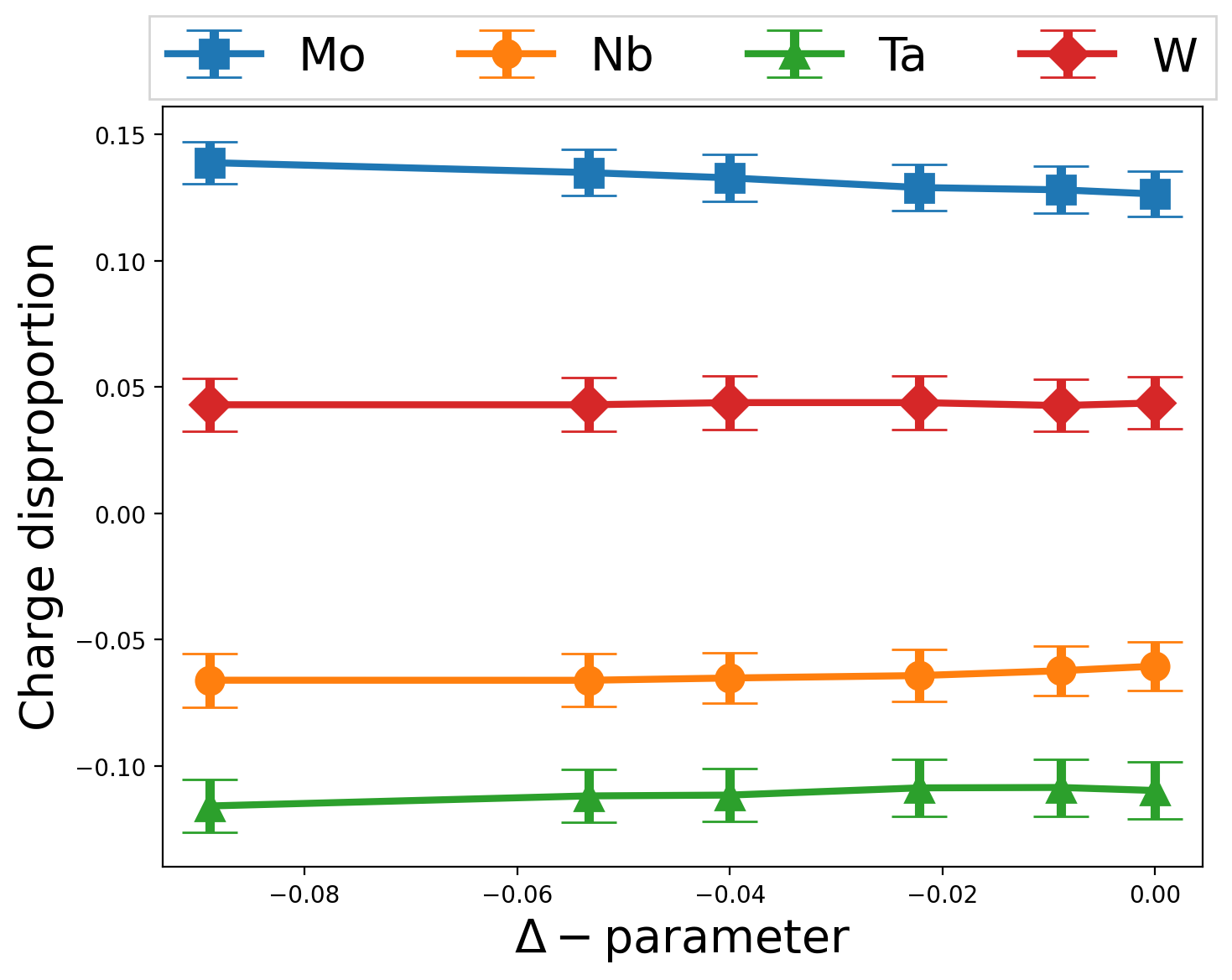}
        \caption{}
        \label{subfig:mntw_charge}
    \end{subfigure}
    \begin{subfigure}[b]{0.3\textwidth}
        \centering
        \includegraphics[width=\textwidth]{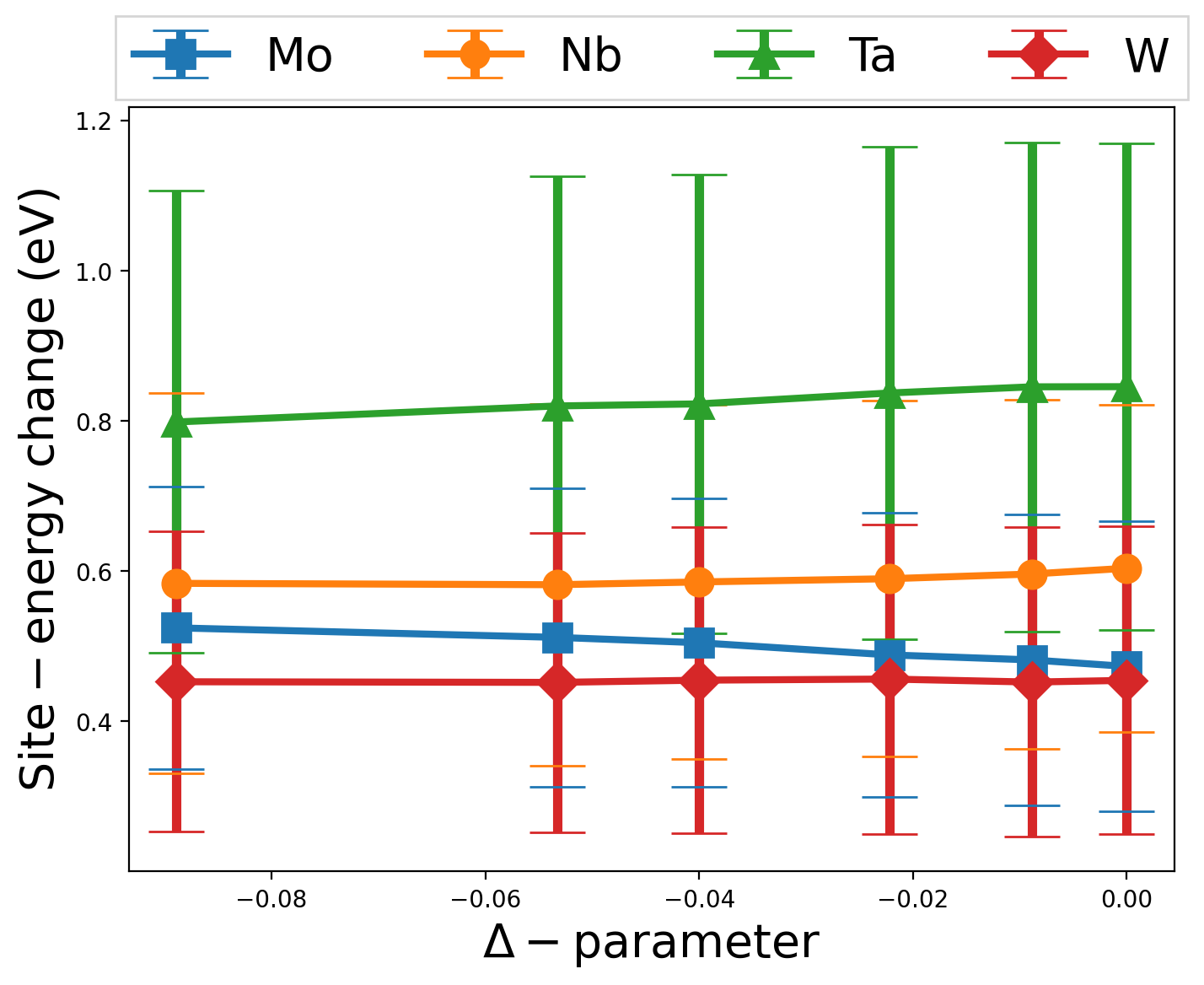}
        \caption{}
        \label{subfig:mntw_siteene}
    \end{subfigure}
    \caption{The variation in the (\ref{subfig:mntw_charge}) charge disproportion, and (\ref{subfig:mntw_siteene}) atomic site-energies for MoNbTaW alloy with Mo-Ta ordering. Note that error bars represents 95\% interval here.}
    \label{fig:mntw_analysis}
\end{figure*}

We validate the proposed order-parameter or $\mathrm{\Delta-parameter}$ with respect to the energetic change associated with the occurrence of the order in FCC medium entropy alloy (CoCrNi) and BCC (MoNbTaW) high-entropy alloy (Fig \ref{fig:cr_cr} and \ref{fig:mo_ta}). We use first-principles based linear scaling Density Functional Theory code, Locally Self-consistent Multiple Scattering (LSMS) \cite{LSMS1,LSMS2} to calculate the energies, magnetic moments and charges in these systems. Within LSMS, we use all parameters such that the energies are converged to within tolerances of $10^{-6}$ Ry/atom. We introduce the order in CoCrNi system by increasing the propensity of Cr-Cr bonds (or causing the increase in the $\mathrm{\Delta-parameter}$) and studied the energy difference due to such variation. It can be seen that increased Cr-Cr interactions lead to the greater energy value, which has been reported in the literature \cite{xu2022critical}, while for BCC high entropy alloy, we order the Mo-Ta bond and it can be seen that with increase in the number of Mo-Ta bond and hence decrease in the $\mathrm{\Delta}$ value, the energy of the alloy decreases, which again agrees with the reported results \cite{Widom2014}. We also studied the effect of the ordering on the charge-disproportion (\emph{i.e.,} the difference in the charge on the atom, before and after the electronic minimisation. The negative value implies the gain in charge and \emph{vice-versa}), atomic magnetic moments and site-energies of atoms. It can be seen that the statistically charge-disproportion and site-energies for the Co and Ni atom shows the decreasing trend, while for the Co atoms, opposite trend can be seen (Fig. \ref{fig:ccn_charge} and \ref{fig:ccn_siteene}). The magnetic moment for Co and Ni show weakly increasing trend, while opposite can be seen for Cr (Fig. \ref{fig:ccn_magmom}).In the case of Mo-Ta bond ordering in MoNbTaW alloy, it can be seen that with increase in the ordering, the charge disproportion and site-energies show decrement for Mo atoms, while increment for Nb and Ta atoms. It is interesting to see that W atoms do not show any clear trend (Fig. \ref{fig:mntw_analysis}). It should additionally be noted that with simply variation in the order parameter by $~\mathrm{10^{-2}}$, we can observe noticeable effect on the charge transfer characteristics, which points towards the  long-ranged effect in charge transfer in high-entropy alloys \cite{karabin2022ab}.\\

\section*{Case of ordering in high-entropy oxide} 
\begin{figure*}[!htb]
    \centering
    \includegraphics[width=0.8\textwidth]{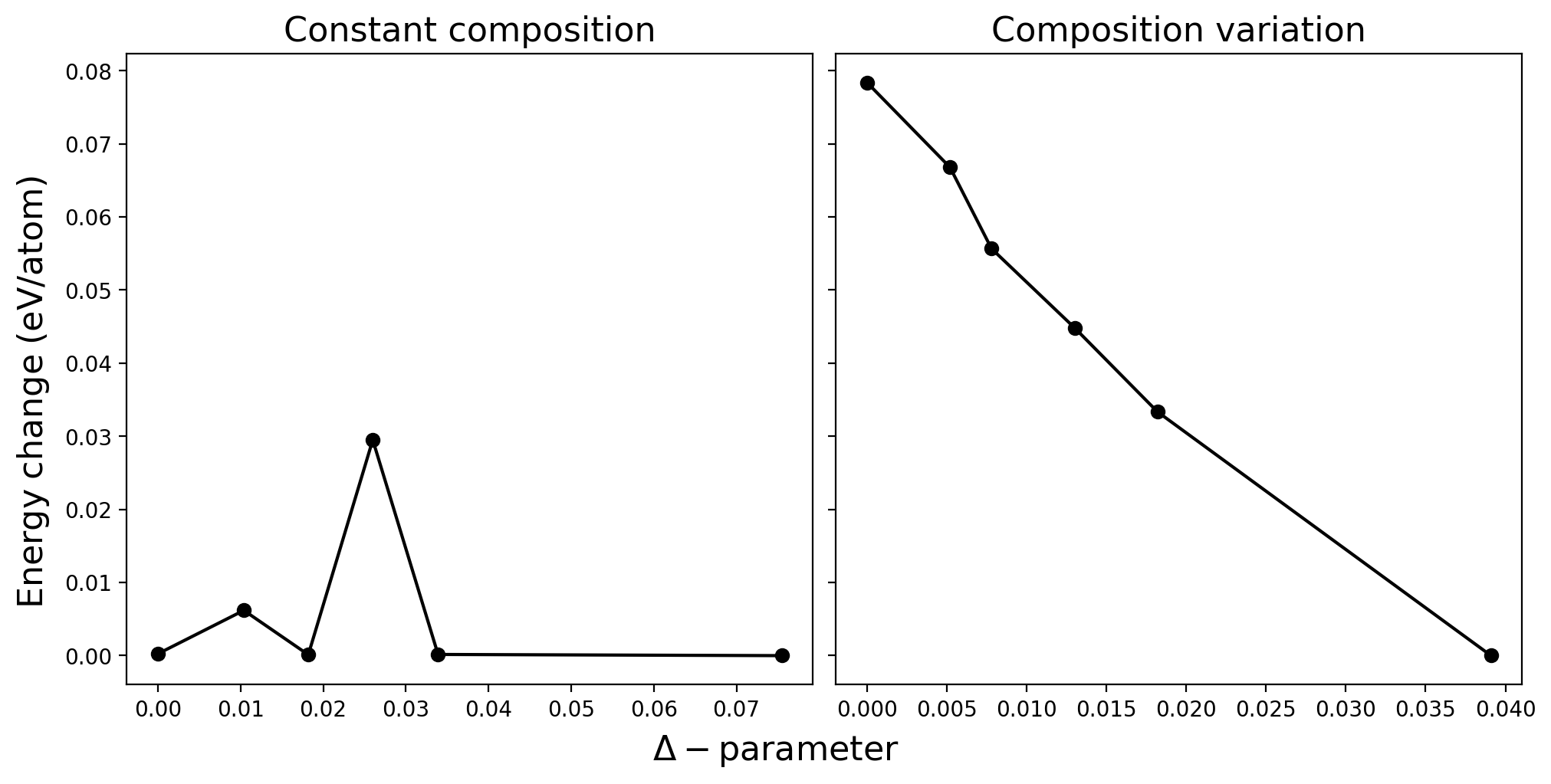}
    \caption{The variation in the energy of the (CoCuMgNiZn)O high-entropy oxide with variation in the $\mathrm{Cu^{2+}-O^{2-}}$ bond number, while keeping the constant composition and in the second case, increasing the $\mathrm{Cu^{2+}-O^{2-}}$ bond number (and corresponding decrease in $\mathrm{Co^{2+}-O^{2-}}$ bond number) by varying the composition.}
    \label{fig:heo_ene}
\end{figure*}

We additionally employ the OPERA framework to generate the cation ordering in (CoCuMgNiZn)O, high-entropy oxide. In such cases, the ordering can take place on multiple sub-lattices. Figure \ref{fig:heo_ene} shows the variation in the energy with the $\mathrm{\Delta}$ order-parameter for two cases. In the first case (constant composition), the composition of the supercell is kept constant and $\mathrm{Cu^{2+}-Cu^{2+}}$ instances are increased, as carried out for alloys in earlier section. In this case, the ordering was being carried out only on the cation sub-lattice (\emph{i.e.,} second nearest-neighbour ordering). It can be seen that with increase in the ordering does not lead to trend in the energy. Such observation can be rationalised in terms of the observation that, since anion sub-lattice has only $\mathrm{O}$ anion, the effect of first nearest-neighbour bond energetics do not influence the total energy of the system. But, when we vary the composition of the supercell to increase the number of $\mathrm{Cu^{2+}-O^{2-}}$ bonds by decreasing the $\mathrm{Co^{2+}-O^{2-}}$ bond, the energy of the system decreases with a trend, as observed earlier for such oxides \cite{anand2018phase}. Through this approach, we demonstrate that OPERA approach can be employed for varying composition of the supercell in desried manner. \\
It should also be added that in the present investigation the applicability of the OPERA framework for the single site ordering with ordering taking place in the cation sub-lattice has been demonstrated. However, the OPERA framework can be applied to study the complex ordering at multiple sub-lattices, which can take place in other high-entropy systems such as perovskites \cite{jiang2018new,halder2021understanding}, carbonitrides \cite{dippo2020bulk, wen2020thermophysical}, boronitrides \cite{qin2020dual},  and other complex ceramics \cite{mccormack2021thermodynamics, toher2022high}.  \\        

\section*{Conclusions}
The proposed work provides a high-throughput framework or OPERA (\underline{O}rder \underline{P}arameter \underline{E}ngineering of \underline{RA}ndom Systems) scheme for studying the chemical short-range order in high-entropy materials. Instead of generating the CSRO configurations through computationally expensive scheme, which requires either \emph{ab-initio} calculations or interatomic force-fields, the OPERA method generates the desired configurations through combinatoric sampling. Such method allows for the exploration of not only equilibrium, but also non-equilibrium structures. This framework inherits the merit of Warren-Cowley order parameter (\emph{i.e.,} different order parameter for clustering of like-atoms and superstructure formation due to the unlike-atoms). But it does not suffer from the fundamental limitation of the Warren-Cowley parameter, where the composition of the configuration need to be changed (where system needs to be in contact of infinite reservoir of atoms) and order parameter for each pair of atoms need to be defined. In contrast to the above, the proposed $\mathrm{\Delta}$-parameter in the OPERA scheme can generate the desired CSRO trend as well as it can be quantified as single scalar value or $\mathrm{\Delta}$-value.   
\section*{Code availability}
The OPERA code and representative input files can be assessed at \url{https://github.com/ganand1990/OPERA}. 
\section*{Acknowledgment}
GA is thankful to National Param Supercomputing Facility for providing access to the Param-Yuva II facility. This research used resources of the Oak Ridge Leadership Computing Facility, which is supported by the Office of Science of the U.S. Department of Energy under Contract No. DE-AC05-00OR22725.

\bibliography{manuscript}






\begin{center}
    \section*{\LARGE{Supplementary Information}}
\end{center}
\section{Number of like and unlike bonds in chemically-disordered supercell with compositional constraint}\label{sec:A1}
\begin{figure*}[!htb]
    \centering
    \includegraphics[width=0.6\textwidth]{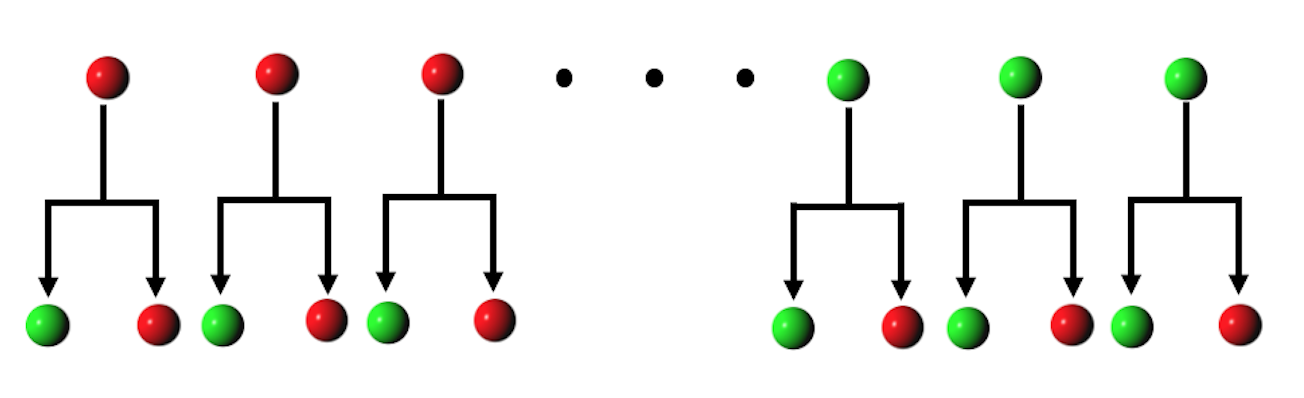}
    \caption{Schematic showing the bond-counting method for a hypothetical binary system containing 2 atoms as the nearest-neighbours.}
    \label{fig:random_count}
\end{figure*}
Consider a chemically disordered supercell containing atoms A and B in the equal proportion. In the perfectly disordered system, the $\mathrm{i^{th}}$-coordination shell should contain equal number of A and B atoms, irrespective of the atom-type in the centre. If atom `A' is in the centre, equal number of A and B atoms will be present in the $\mathrm{i^{th}}$ coordination shell.\\
The bond-counting as shown in Fig. \ref{fig:random_count} for hypothetical binary system with 2 nearest-neighbour. Note that as the bonds are counted for each atom, the like-bond occurrence can only occur for the same atom, while the unlike-bond can occur for centre-atom of its type as well as for the other type centre-atom. Now, if $\mathrm{N}$ is the total number of atoms and A and B are present in equal proportion. Then, for $\mathrm{\left(\frac{N}{2}\right)}$ A centre-atoms, number of $\mathrm{A-A}$ bond $\mathrm{\left(N_{AA}\right)}$ is $\mathrm{\left(\left(\frac{N}{2}\right) \cdot n_i \cdot P(A|A) \right)}$, where $\mathrm{n_i}$ is the number of atoms in the $\mathrm{i^{th}}$ coordination shell and  number of $\mathrm{A-B}$ bonds $\mathrm{\left(N_{AB}\right)}$ is  $\mathrm{\left(\left(\frac{N}{2}\right) \cdot n_i \cdot P(A|B) \right)}$. So, total number of bonds for A-centred atoms and B-centred atoms is,
\begin{equation}
    \mathrm{
    N_{AA} + N_{AB} + N_{BB} + N_{BA} = \left[\left(\frac{N}{2}\right) \cdot n_i  \right] \cdot \left[P(A|A) + P(A|B) + P(B|B) + P(B|A) \right]
    }
\end{equation}
Since, $\mathrm{N_{AB} = N_{BA}}$, there is the double counting for the unlike-atoms. Hence, due to the compositional constraint of keeping the composition of the supercell constant, the number of unlike-bonds in the chemically disordered supercell is twice to that like-bonds.

\section{Introduction of CSRO without any constraints}\label{sec:A2}
\paragraph{Case-1: $\mathrm{\Delta > 0}$}
As mentioned above, Positive CSRO can be imposed on the system by increasing the instances of a particular like-bond instance (preferred like-bond). If a particular preferred like-bond instance needs to be increased, with corresponding decrease in the other non-preferred bonds (as the number of bonds in the supercell is constant), the equation \ref{one_eq} may be rewritten as,
\begin{equation}\label{third}
\mathrm{
    \Delta=\frac{K_1 - K_1\cdot\left(\frac{2n-a_1}{2n}\right)}{K_1} + \frac{\left(K_2-1\right)\cdot\left(\frac{n-a_2}{n}\right)+ \left(\frac{n+a_3}{n} \right) - K_2}{K_2}
    }
\end{equation}
Above expression can be understood as; the imposition of positive CSRO implies the deviation from the perfect chemical disorder, where $\mathrm{m_{ij}=2n}$ and $\mathrm{m_{ii}=n}$. $\mathrm{a_1}$, $\mathrm{a_2}$ and $\mathrm{a_3}$ represent the reduction in the non-preferred unlike bonds, reduction in the non-preferred like bonds and increase in the preferred like bonds instances, respectively. Rearranging the equation \ref{third}, we have,
\begin{equation}
\mathrm{
    \frac{K_2\cdot a_1}{2n}-\frac{K_2 \cdot a_2}{n}+\frac{a_2}{n}+\frac{a_3}{n}=K_2\cdot\Delta
    }
\end{equation}
Also, $\mathrm{a_1=2 \cdot a_2}$ and $\mathrm{a_2\cdot \left( 2K_1+K_2-1\right)=a_3}$, as it has been assumed that increase in the preferred bond instance should be compensated by decrease in all the non-preferred bond-instances by similar amount of variation from their value at the complete disorder. So, equation \ref{third} may be further modified as,
\begin{equation}
\mathrm{
    \frac{a_2}{n}+\frac{a_3}{n}=K_2\cdot\Delta
    }
\end{equation}
The final expression for the amount of reduction in the non-preferred bonds may be expressed as,
\begin{equation}
    \mathrm{
    a_1=2\cdot a_2 = \frac{2n\cdot K_2 \cdot \Delta}{2K_1 + K_2}
    }
\end{equation}
Using the relation between $\mathrm{a_2}$ and $\mathrm{a_3}$, we may express the amount of increase in the preferred bond instances, \emph{i.e.,} $\mathrm{a_3}$ as,
\begin{equation}
    \mathrm{
    a = a_3=\frac{2n\cdot K_2 \cdot \Delta \cdot \left( 2K_1+K_2-1\right)}{2K_1 + K_2}
    }
\end{equation}
\paragraph{Case-2: $\mathrm{\Delta < 0}$} 
Equation \ref{one_eq} is modified to increase the preferred unlike bond instances to impose negative $\mathrm{\Delta}$ and it may be represented as,
\begin{equation}
    \mathrm{
    \Delta=\frac{K_1 - \left(K_1 -1 \right) \cdot \left(\frac{2n-a_1}{2n} \right) - \left(\frac{2n+a_2}{2n} \right)}{K_1} + \frac{K_2 \cdot \left(\frac{n-a_3}{n}\right) - K_2}{K_2}
    }
\end{equation}
Above may be simplified as,
\begin{equation}
    \mathrm{
    \frac{K_1 a_1}{2n}-\frac{a_1}{2n}-\frac{a_2}{2n}-\frac{K_1 a_3}{n}=K_1 \cdot \Delta 
    }
\end{equation}
Since, $\mathrm{a_1 = 2 \cdot a_3}$ and $\mathrm{a_2 = a_3\left(2K_1 + K_2 - 2\right)}$, above would yield, 
\begin{equation}
    \mathrm{
    a_1=2a_3=-\frac{4n \cdot \Delta \cdot K_1}{2K_1 + K_2}
    }
\end{equation}
and,
\begin{equation}
    \mathrm{
    a = a_2=-\frac{2n \cdot \Delta \cdot K_1 \cdot \left(2K_1+K_2-2 \right)}{2K_1 + K_2}
    }
\end{equation}
Note that negative sign gets cancelled with the negative sign of the $\mathrm{\Delta-parameter}$ to provide positive values of $\mathrm{a_1}$, $\mathrm{a_2}$ and $\mathrm{a_3}$.

\section{Introduction of CSRO with atomic identity constraint}
Consider a scenario, where the bond propensity for certain bond needs to be increased, while keeping the composition of the supercell constant. It should be noted that if the particular bond instance need to be generated by purely a swap process, then certain bonds among all the possible like and unlike bonds can be swapped to generate the desired bond. Considering the two cases again for generating the configurations with negative as well as positive values of the $\mathrm{\Delta}$-parameter.\\
\paragraph{Case-1: $\mathrm{\Delta < 0}$}
The generation of the preferred unlike-bond instance can take place by the swap of the bonds, which at least have one of the element of the preferred unlike bond. This may be understood from giving an example of a system containing 5 elements, ABCDE. Let us consider that the preferred unlike bond is A-B. In the suggested system, A-A, A-B, A-C, A-D, A-E, B-B, B-C, B-D, B-E, C-C, C-D, C-E, D-D, D-E and E-E bonds are present. If, we want to increase the propensity of A-B bond, then the swap between [A-A, A-C, A-D, A-E] and [B-B, B-C, B-D, B-E] needs to be carried out. So, for an increase in the pair of preferred unlike-bond instance (A-B), there is a reduction of a pair of like-bond instance (A-A and B-B, if the swap between them takes place) and decrease in other unlike-bond instances. 
\begin{table}[h!]
    \centering
    \begin{tabular}{ | p {1.5cm} | p {2.5cm} | p {2.5 cm} | p {2.5cm} | p {2.5cm}}
    \hline
         e & Decrease in like-bond & Increase in like-bond & Decrease in unlike-bond & Increase in unlike-bond 
         \\
         \hline
        2 & 2 &  0 & 2 &  0 \\
         3 &  2 &  1 &  3 &  2 \\
        4 & 2 &  2 & 4 &  4 \\
        5 & 2 & 3 &  5 &  6 \\
         6 &  2 &  4 &  6 &  8 \\
        \hline
         \textbf{\tiny{Analytical expression} (e)} &  2 & e-2 &  e & 2(e-2) \\
        \hline
     \end{tabular}
    \caption{}
    \caption{With variation in the number of type of elements in the system, table shows the variation in the like and unlike bonds, when the $\mathrm{\Delta < 0}$ is desired through increasing the propensity of the unlike-bond.}
    \label{tab:a3}
\end{table}
Table \ref{tab:a3} shows the variation in the type of bond with the swap process. Say, for a five-component alloy, swap between [A-A, A-C, A-D, A-E] and [B-B, B-C, B-D, B-E] takes place. It is clear that such swap would lead to the reduction of 2 like-bonds (\emph{i.e.,} A-A and B-B) and 6 unlike-bonds (\emph{i.e.,} A-C, A-D, A-E, B-C, B-D and B-E). While, there is increase of 3 like-bonds (C-C, D-D and E-E, due to the swap of (A-C, B-C), (A-D, B-D) and (A-E, B-E), respectively) and 6 instances of preferred A-B bond is generated for the one iteration of the swap.\\
The equation \ref{one_eq}  can now be modified as,
\begin{equation}
    \centering
    \mathrm{
    \Delta = \frac{\sum\limits^{e}{1 - \left(\frac{2n + a_1}{2n} \right)} + \sum\limits^{2(e-2)}{1 - \left(\frac{2n - a_2}{2n} \right)}}{K_1} + \frac{\sum\limits^{2}{\left(\frac{n - a_3}{n} \right) - 1} + \sum\limits^{e-2}{\left(\frac{n + a_4}{n} \right)-1}}{K_2}
    }
\end{equation}
Note that other terms of $\mathrm{\Delta}$-parameter, which are not varied goes to zero. Now, if $\mathrm{a_3 = a_4 = \frac{a_4}{2}}$ and $\mathrm{e \cdot a_1 = -2(e-2) \cdot a_2 - 2a_3 + (e-2)\cdot a_4}$, we have,
\begin{equation}
    \centering
    \mathrm{
    a_1 = \frac{-5a_2}{2} - \frac{2a_2}{e}
    }
\end{equation}
Using above, the value of $\mathrm{a_1}$ may be expressed as,
\begin{equation}
    \centering
    \mathrm{
    a = a_1 = \frac{\Delta \cdot K_1 \cdot K_2 \cdot \left(-5e-4 \right)}{7e^2 \cdot K_2 + e^2 \cdot K_1 - 4e \cdot K_1}
    }
\end{equation}
\paragraph{Case-2: $\mathrm{\Delta > 0}$}
The configuration with the positive value of the $\mathrm{\Delta}$-parameter can be generated with the increase in the like-preferred bond instances. Considering the case of five-component (ABCDE) alloy again. Let us consider that bond A-A needs to be increased. The A-A bond can be increased only with swap among A-B, A-C, A-D and A-E. By the swap their is decrease in unlike bond instances, while there is increase in the preferred like bond instances. \\
\begin{table*}[h!]
    \centering
    \begin{tabular}{| p {1.5cm} | p {3.5cm} | p {3.5 cm} | p {3.5cm}|}
    \hline
         e & Increase in like-bond &  Decrease in unlike-bond & Increase in unlike-bond
         \\
         \hline
         2 & 0 & 0 & 0 \\
         3 & 1 & 2 & 1 \\
         4 & 1 & 2 & 1 \\
         5 & 2 & 4 & 2 \\
         6 & 2 & 4 & 2 \\
        \hline
        \centering \textbf{\tiny{Analytical expression} (e)} & $\mathrm{\left(e - \floor{\frac{e}{2}} - 1 \right)}$ & $\mathrm{2 \cdot \left(e - \floor{\frac{e}{2}} - 1 \right)}$ & $\mathrm{\left(e - \floor{\frac{e}{2}} - 1 \right)}$ \\
        \hline
    \end{tabular}
    \caption{variation in the number of type of elements in the system, table shows the variation in the like and unlike bonds, when the $\mathrm{\Delta > 0}$ is desired through increasing the propensity of the like-bond. The $\mathrm{\floor{}}$ in the table represents the \emph{floor} operator.}
    \label{tab:a4}
\end{table*}
The equation \ref{one_eq} for this particular case may be written as,
\begin{equation}
    \centering
    \mathrm{
    \Delta = \frac{\sum\limits^{2 \cdot \left(e - \floor{\frac{e}{2}} - 1 \right)}{\left(1 - \frac{2n - a_1}{2n}\right)} + \sum\limits^{\left(e - \floor{\frac{e}{2}} - 1 \right)}{\left(1 - \frac{2n + a_2}{2n}\right)}}{K_1} + 
    \frac{\sum\limits^{\left(e - \floor{\frac{e}{2}} - 1 \right)}{\left(\frac{n + a_3}{n} - 1 \right)}}{K_2}
    }
\end{equation}
If, $\mathrm{a_1 = a_2}$ and $\mathrm{-2 \cdot \left(e - \floor{\frac{e}{2}} - 1 \right) \cdot a_1 + \left(e - \floor{\frac{e}{2}} - 1 \right) \cdot a_2 = \left(e - \floor{\frac{e}{2}} - 1 \right) \cdot a_3}$, then above may be simplified as,
\begin{equation}
    \centering
    \mathrm{
    a = a_3 = \frac{2 \cdot \Delta \cdot n \cdot K_1 \cdot K_2}{\left(2K_1 - K_2 \right) \cdot \left(e - \floor{\frac{e}{2}} - 1 \right)}
    }
\end{equation}


\end{document}